\documentclass[pdflatex,sn-apa]{sn-jnl}

\usepackage{amsthm,amsfonts,amsmath,amssymb,stmaryrd}
\usepackage{hyperref,graphicx,tikz}
\usepackage{mathtools}
\usepackage{xurl}

\usepackage{enumitem}
\setlist[enumerate]{label=(\arabic*),ref=\arabic*}
\usetikzlibrary{arrows.meta, positioning,calc,decorations.pathreplacing}
\theoremstyle{thmstylethree}
\newtheorem{definition}{Definition}

\theoremstyle{thmstyleone}
\newtheorem{proposition}[definition]{Proposition}
\newtheorem{lemma}[definition]{Lemma}
\newtheorem{theorem}[definition]{Theorem}
\newtheorem{corollary}[definition]{Corollary}

\newtheorem*{claim*}{Claim}

\theoremstyle{thmstyletwo}

\title{Properties of Public Announcements as Operators: Commutativity, Monotonicity, Idempotence, Fixed Points, and Stabilization}
\date{\today}
\author*[1]{\fnm{Eiji} \sur{Yamada}}\email{yamada.e.112d@m.isct.ac.jp}
\affil*[1]{\orgdiv{Department of Mathematical and Computing Science}, \orgname{Institute of Science Tokyo}, \state{Tokyo}, \country{Japan}}

\abstract{For each formula $\varphi$, we introduce the public announcement operator $U_\varphi$, which maps every model to its updated model under the public announcement of $\varphi$. We then study its algebraic and dynamic properties with a particular focus on commutativity, monotonicity, idempotence, fixed points, and stabilization. Among many results, we show the following. First, monotonicity of $U_\varphi$ is characterized by a family of preservation conditions and equivalence to an existential formula. 
Second, there exists a formula that achieves every nonzero ordinal as a stabilization rank---the first stage at which a model stabilizes---of some model. Finally and unexpectedly, extinction of every model under iterated announcements of $\varphi$ is equivalent to common unknowability, being always informative when true, and a fixed point view of the Moore sentence. Over S5, those are further equivalent to common unbelievability and being always informative. Other results include characterizations of commutativity and strong commutativity, idempotence as success, properties of fixed points, and finite and $\omega$-stabilization.
For analysis, we use PAL (state elimination) and BPAL (arrow elimination), clarifying their subtle differences.}


\begin{document}

\maketitle

\section{Introduction}
Epistemic logic is the logical study of knowledge and belief, and dynamic epistemic logic is the logical study of information flow, knowledge acquisition, and belief revision~\citep{faginEtAl1995,vanDitmarschHoekKooi2007}. \emph{Public announcements} are one of the main topics in the field. Here, a public announcement of $\varphi$ is an announcement to a group of agents that $\varphi$ holds.

\emph{Public announcement logic} (PAL) was proposed by \citet{plaza2007} to express this action. Announcing $\varphi$ deletes the states at which $\varphi$ is false in the current model. That is, for a (Kripke) model $M=(W,\{R_i\}_{i\in G},V)$, the updated model $M_{|\varphi}=(W_{|\varphi}, \{R_{i|\varphi}\}_{i\in G},V_{|\varphi})$ in PAL is defined by $W_{|\varphi}=\llbracket\varphi\rrbracket_M$, $R_{i|\varphi}=R_i\cap (W_{|\varphi}\times W_{|\varphi})$ for each agent $i\in G$, and $V_{|\varphi}(p)=V(p)\cap W_{|\varphi}$ for all proposition letters $p$. Here, $\llbracket\varphi\rrbracket_M$ is the set of states in $M$ in which $\varphi$ holds.
On the other hand, \emph{believed public announcement logic} (BPAL) was also proposed~\citep{GerbrandyGroeneveld1997,agotnesVanDitmarschWang2018}, which deletes all arrows of accessibility relations that go to states where $\varphi$ is false. That is, the updated model $M|\varphi=(W,\{R_i|\varphi\}, V)$ in BPAL is defined by $w(R_i|\varphi) v$ iff $wR_iv$ and $M,v\models\varphi$ for all agents $i\in G$ and states $w,v\in W$. 

Thus, PAL is \emph{state elimination} and BPAL is \emph{arrow elimination}. Therefore, BPAL can express not only truthful announcements but also false announcements: In PAL, we can no longer consider the truth of $\varphi$ at a state $w$ after a public announcement of $\varphi$ whenever $\varphi$ is initially false, simply because $w$ gets eliminated after an announcement. In contrast, in BPAL we can always consider the truth of $\varphi$ regardless of the initial truth value. 

In this paper, we introduce, for each formula $\varphi$, the \emph{public announcement operator} $U_\varphi$ defined as $U_\varphi(M)=M_{|\varphi}$ in PAL and $U_\varphi(M)=M|\varphi$ in BPAL for all models $M$. We then study its algebraic properties and dynamic properties with a particular focus on commutativity ($U_\varphi\circ U_\psi=U_\psi\circ U_\varphi$), monotonicity ($M\preceq N\Rightarrow U_\varphi(M)\preceq U_\varphi(N)$), idempotence ($U_\varphi^2=U_\varphi$), fixed points ($U_\varphi(M)=M$), and stabilization ($ U_\varphi^\alpha(M)=U_\varphi^{\alpha+1}(M)$ for some ordinal $\alpha$). More specifically, algebraic properties studied in this paper are associativity, identity, zero elements, inverse elements, commutativity, strong commutativity ($U_\varphi\circ U_\psi=U_{\varphi\land\psi}=U_\psi\circ U_\varphi$), idempotence ($U_\varphi^2=U_\varphi$), left/right absorption, and deflationarity ($U_\varphi(M)\preceq M$) as summarized in Table~\ref{tab:algebraic-properties-announcement-operators} of Section~\ref{sec:algebraic-properties}. Dynamic properties studied in this paper are fixed points, stabilization, and stabilization rank as summarized in Table~\ref{tab:announcement-operator-properties} of Section~\ref{sec:dynamic-properties}. 

Our results include the following. For algebraic properties, Theorems~\ref{thm:commutativity} and \ref{thm:strong-commutativity} give equivalent conditions for commutativity and strong commutativity of $U_\varphi$, respectively. Corollary~\ref{cor:commutativity-and-strong-commutativity} gives the difference between commutativity and strong commutativity. Theorem~\ref{thm:monotonicity} gives equivalent conditions for the monotonicity of $U_{\lnot\varphi}$. These include a family of preservation properties of $\varphi$, such as remaining true when states or accessibility arrows are deleted, equivalence to a universal formula, and two further conditions involving public announcements. Monotonicity of $U_\varphi$ is then equivalent to their dual conditions, including equivalence to existential formulas. Proposition~\ref{prop:deflationarity-and-idempotence} states deflationarity of $U_\varphi$ for any formula $\varphi$, and identifies idempotence of $U_\varphi$ with \emph{success} and the condition $U_\varphi^2=U_\bot$ with \emph{self-refutation}~\citep{vanDitmarschKooi2006,hollidayIcard2010}.

For dynamic properties, Theorem~\ref{thm:fixed-points} gives a characterization of fixed points, the set $\operatorname{Fix}(U_\varphi)$ of fixed points of $U_\varphi$, and special properties of $U_\varphi$ and $U_\varphi^\infty$ which hold when $\varphi$ is either universal or existential. We also define 
the \emph{stabilization rank} $\widehat{\rho}_\varphi(M)$ of a model $M$ under iterated announcements of $\varphi$ to be the least ordinal $\alpha$ such that $U_\varphi^\alpha(M)=U_\varphi^{\alpha+1}(M)$. We also use $\widehat{\rho}_\varphi^P$ for PAL and $\widehat{\rho}_\varphi^B$ for BPAL. 
Theorem~\ref{thm:stabilization-rank} then gives properties of stabilization rank. In particular, Theorem~\ref{thm:stabilization-rank}(4) states that there exists a formula $\varphi$ such that, for each ordinal $\alpha>0$, there is a model $M_\alpha$ satisfying $\widehat{\rho}_\varphi^P(M_\alpha)=\widehat{\rho}_\varphi^B(M_\alpha)=\alpha$. Furthermore, this $\varphi$ can be taken to be \emph{always informative when true} (i.e., for all pointed models $M,w$, $M,w\models\varphi$ implies $U_\varphi(M)\neq M$)~\citep{hollidayIcard2010,yamada2026Eventual}. Therefore, Theorem~\ref{thm:stabilization-rank}(4) implies that \emph{being always informative when true does not imply fast stabilization of any model} (i.e., there need not be a uniform bound on stabilization rank). Theorems~\ref{thm:finite-stabilization} and \ref{thm:omega-stabilization} give equivalent conditions for finite stabilization ($\widehat{\rho}_\varphi< \omega$) and $\omega$-stabilization ($\widehat{\rho}_\varphi\leq \omega$). According to the former theorem, if every model stabilizes within finite time under iterated announcements of $\varphi$, there actually exists a uniform finite bound $N$ on the stabilization rank $\widehat{\rho}_\varphi$. 

Finally, we ask exactly when iterated announcements
of $\varphi$ make a model empty or edgeless (model extinction). We also introduce the notions of common unknowability/unbelievability. We say that $\varphi$ is \emph{commonly unbelievable} if $C_G\varphi$ is unsatisfiable and that $\varphi$ is \emph{commonly unknowable} if $\varphi\land C_G\varphi$ is unsatisfiable, following the notions of unbelievability and unknowability of $\varphi$ in \citet{yamada2026Sources}. Unexpectedly, Theorem~\ref{thm:extinction-under-iterated-announcements} states that model extinction is equivalent
to common unknowability and always informativeness when true, and hence to a fixed point
view of the generalized Moore sentence $\varphi\land\lnot C_G\varphi$ in \citet[Theorem~13]{yamada2026Eventual}. Over S5, those are further equivalent to common unbelievability and always informativeness (both when true and when false).
We consider Theorems~\ref{thm:monotonicity}, \ref{thm:stabilization-rank}, and \ref{thm:extinction-under-iterated-announcements} of particular interest. 

For related work, \citet{vanDitmarschKooi2006} investigate success. \citet{hollidayIcard2010} analyze success and self-refutation. They also introduce eventual self-refutation and always informativeness (when true, in our terminology), and prove their equivalence with the unsatisfiability of $\Box\varphi$ in single-agent S5. \citet{yamada2026Sources} introduces unknowability and unbelievability, and shows that all the sources of unknowability are ``Moorean phenomena.'' \citet{agotnesVanDitmarschWang2018} study success, self-refutation,
true lies, and impossible lies, as well as truth-value patterns under iterated announcements. \citet{yamada2026Classification} then resolves their conjecture on the classification of truth patterns in terms of $\sigma$-valid formulas. \citet{yamada2026Eventual} introduces eventual and strong eventual versions of the four notions, as well as their transfinite versions. He then analyzes their logical relationship.

There is a dynamical-systems perspective on iterated \emph{action model} updates. Here, \emph{action model logic}~\citep{baltagMossSolecki2016} is a logic that can express not only our PAL and BPAL updates but also more general epistemic actions including private announcements or even nondeterministic updates. An action model is a structure in that logic. \citet{Sadzik2006} analyzes iterated action-model updates and their
asymptotic stability. \citet{KleinRendsvig2020}
study convergence, continuity, and recurrence on spaces of pointed models modulo logical
equivalence, equipped with topologies appropriate to logical convergence.

A complementary line of research incorporates iteration into the object language. \citet{millerMoss2005} prove undecidability results for languages with iterated
modal relativization. \citet{liWang2026} give a sound and
complete axiomatization and a decidability result for a single-agent S5 extension of PAL
with tests, choice, sequential composition, and Kleene-star iteration.

The treatment of fixed points and transfinite iteration has 
precedents in epistemic game theory. \Citet{vanBenthem2007} relates iterated
public announcements to rationality and iterative strategy elimination. \citet{aptZvesper2007,aptZvesper2010} study arbitrary strategy sets, common
beliefs, and the role of monotonicity in such elimination procedures, including
transfinite iterations. \citet{masuzawaHasebe2011} give conditions
under which iterative information updates lead to stable outcomes independently of the
order of updating.

This paper is structured as follows. Section~\ref{sec:basic-definitions} gives basic definitions including PAL, BPAL, the four notions, and the public announcement operators. Sections~\ref{sec:algebraic-properties} and \ref{sec:dynamic-properties} investigate algebraic properties and dynamic properties, respectively. Section~\ref{sec:conclusion} reviews the results and discusses future directions.

A Lean formalization is available at \url{https://github.com/eiyamada/lean-paper-formalizations}.

\section{Basic definitions}\label{sec:basic-definitions}
We first list several basic definitions. Let $G$ be a nonempty finite set of agents and $\mathbf{Prop}$ be a countably infinite set of proposition letters.
\begin{definition}\label{def:epistemic-language}
  Define \textit{formulas} in multi-agent epistemic logic $\mathcal{L}$ by
  \begin{equation*}
    \varphi::=p\mid\lnot\varphi\mid\varphi\land\psi\mid\Box_i\varphi\quad(p\in \mathbf{Prop},\quad i\in G)
  \end{equation*}
\end{definition}
\begin{definition}[\citealp{plaza2007}]\label{def:public-announcement-language}
  Define \textit{formulas} in public announcement logic $\mathcal{L}_{PAL}$ by
  \begin{equation*}
    \varphi::=p\mid\lnot\varphi\mid \varphi\land\psi\mid \Box_i\varphi\mid [!\varphi]\psi\quad (p\in\mathbf{Prop},\quad i\in G)
  \end{equation*}
\end{definition}

\begin{definition}[\citealp{agotnesVanDitmarschWang2018}]\label{def:believed-public-announcement-language}
  Define \textit{formulas} in believed public announcement logic $\mathcal{L}_{\text{BPAL}}$ by
  \begin{equation*}
    \varphi::=p\mid\lnot\varphi\mid \varphi\land\psi\mid \Box_i\varphi\mid [\uparrow\varphi]\psi\quad (p\in\mathbf{Prop},\quad i\in G)
  \end{equation*}
\end{definition}
Other propositional connectives are understood as abbreviations and we define duals by $\Diamond_i\varphi:=\lnot\Box_i\lnot\varphi$, $\langle ! \varphi\rangle\psi:=\lnot[!\varphi]\lnot\psi$, and $\langle \uparrow \varphi\rangle\psi:=\lnot[\uparrow\varphi]\lnot\psi$. Note that $\langle!\varphi\rangle\psi$ is equivalent to $\varphi\land[!\varphi]\psi$. $\langle \uparrow \varphi\rangle\psi$ is equivalent to $[\uparrow\varphi]\psi$ by its semantics so there is no difference between $[\uparrow\varphi]\psi$ and its dual. When referring to both $[!\varphi]$ and $[\uparrow\varphi]$, we use $[\varphi]$ as a convention.
The expressive powers of the three languages above are known to be equivalent via reduction axioms. For $\mathcal{L}$ and $\mathcal{L}_{PAL}$, see Chapter 8 of \citet{vanDitmarschHoekKooi2007}, and for $\mathcal{L}$ and $\mathcal{L}_{BPAL}$, see Section 2 of \citet{agotnesVanDitmarschWang2018}.
\begin{definition}\label{def:models}
  A \textit{model} is a tuple $M=(W,\{R_i\}_{i\in G},V)$ where $W$ is a non-empty set of \textit{states}, for each $i\in G$, $R_i$ is a binary relation on $W$ (\textit{accessibility relation}), and $V\colon \mathbf{Prop}\to 2^W$ is a \textit{valuation function}. A \emph{pointed model} is a tuple $M,w$ with $w$ a state in $M$.
\end{definition}

Let $R^*$ be the reflexive transitive closure of a binary relation $R$.
\begin{definition}\label{def:submodels}
  A model $M'=(W',\{R_i'\}_{i\in G},V')$ is a \textit{submodel} of $M=(W,\{R_i\}_{i\in G},V)$ (written $M'\preceq_P M$) iff $W'\subseteq W$, $R_i'=R_i\cap (W'\times W')$ for all $i\in G$, and $V'(p)=V(p)\cap W'$ for all $p\in\mathbf{Prop}$. The \textit{generated submodel} of $M$  at $w$ is the submodel $M_w=(W_w,\{R_{iw}\}_{i\in G},V_w)$ with $W_w=\{v\in W\colon w\left(\bigcup_{i\in G}R_i\right)^*v\}$.

  For BPAL, define $M'\preceq_B M$ iff $W'\subseteq W$, $R_i'\subseteq R_i\cap(W'\times W')$ for all $i\in G$, and $V'(p)=V(p)\cap W'$ for all $p\in\mathbf{Prop}$.
  Thus $\preceq_P$ is induced-submodel inclusion, whereas $\preceq_B$ also permits deletion of accessibility arrows.
\end{definition}
Throughout the paper, $\preceq$ denotes $\preceq_P$ in PAL and $\preceq_B$ in BPAL. The term \emph{submodel} always refers to an induced submodel, in either logic.

For a formula $\varphi$ and a pointed model $M,w$, let $\llbracket\varphi\rrbracket_M:=\{w\in W\colon M,w\models\varphi\}$ be the set of states in which $\varphi$ holds.
\begin{definition}\label{def:relativized-model}
  Let $M=(W,\{R_i\}_{i\in G},V)$ be a model and $\varphi$ be a formula.
  \begin{itemize}
    \item The \emph{relativization} of $M$ to $\varphi$ under public announcement is the model $M_{|\varphi}=(W_{|\varphi}, \{R_{i|\varphi}\}_{i\in G}, V_{|\varphi})$ where $W_{|\varphi}=\llbracket\varphi\rrbracket_M$, $R_{i|\varphi}=R_i\cap (W_{|\varphi}\times W_{|\varphi})$, and $V_{|\varphi}(p)=V(p)\cap W_{|\varphi}$.
    \item The \emph{relativization} of $M$ to $\varphi$ under believed public announcement is the model $M|\varphi=(W, \{R_i|\varphi\}_{i\in G}, V)$ with $R_i|\varphi=R_i\cap (W\times \llbracket \varphi\rrbracket_M)$.
  \end{itemize}
  We admit the empty model $M=(\varnothing, \{\varnothing\}_{i\in G}, V)$ with $V(p)=\varnothing$ for all $p\in\mathbf{Prop}$ as a convention.
\end{definition}
\begin{definition}\label{def:semantics}
  Let $M=(W,\{R_i\}_{i\in G},V)$ be a model. Define truth as follows.
  \begin{enumerate}
    \item $M,w\models p\iff w\in V(p)$.
    \item $M,w\models\lnot\varphi\iff M,w\not\models\varphi$.
    \item $M,w\models\varphi\land\psi\iff M,w\models\varphi\text{ and }M,w\models\psi$.
    \item $M,w\models\Box_i\varphi\iff M,v\models\varphi$ for all $v$ with $wR_iv$.
    \item $M,w\models[!\varphi]\psi\iff M,w\models\varphi$ implies $M_{|\varphi}, w\models\psi$.
    \item $M,w\models[\uparrow\varphi]\psi\iff M|\varphi, w\models\psi$.
  \end{enumerate}
\end{definition}

We say that a formula $\varphi$ is \emph{valid} in a model $M$ (written $M\models \varphi$) if $M,w\models \varphi$ for all $w\in M$. A \emph{frame} is a tuple $F=(W,\{R_i\}_{i\in G})$, and a model $M=(W,\{R_i\}_{i\in G},V)$ is \emph{based on} $F$.
Throughout the paper, we mainly have the classes of frames K, KD, K45, KD45, and S5 in mind but the classes of frames considered will be specified in each statement. Let $\mathcal{F}$ be a class of frames. We say that $\varphi$ is \emph{satisfiable} in $\mathcal{F}$ if there is a pointed model $M,w$ based on a frame in $\mathcal{F}$ such that $M,w\models\varphi$. Also, $\varphi$ is valid in $\mathcal{F}$ (written $\models_{\mathcal{F}}\varphi$) if $M,w\models\varphi$ for all pointed models $M,w$ based on a frame in $\mathcal{F}$.
\begin{definition}[\citealp{agotnesVanDitmarschWang2018}]\label{def:four-notions}
  Let $\varphi$ be a formula.
    \begin{itemize}
        \item $\varphi$ is \textit{successful} iff $\varphi\to[\varphi]\varphi$ is valid.
        \item $\varphi$ is \textit{self-refuting} iff $\varphi\to[\varphi]\lnot\varphi$ is valid.
        \item $\varphi$ is a \textit{true lie} iff $\lnot\varphi\to[\varphi]\varphi$ is valid.
        \item $\varphi$ is an \textit{impossible lie} iff $\lnot\varphi\to[\varphi]\lnot\varphi$ is valid.
    \end{itemize}
\end{definition}
Define the \emph{public announcement operator} $U_\varphi$ of $\varphi$ as, for each model $M$, $U_\varphi(M)=M_{|\varphi}$ in PAL and $U_\varphi(M)=M|\varphi$ in BPAL.

\section{Algebraic properties}\label{sec:algebraic-properties}
In this section, we investigate the algebraic properties in Table~\ref{tab:algebraic-properties-announcement-operators} with a particular focus on commutativity, strong commutativity, and monotonicity (Section~\ref{subsec:commutativity-strong-commutativity-and-monotonicity}), as well as on deflationarity and idempotence (Section~\ref{subsec:deflationarity-and-idempotence}). 

First note that associativity and the existence of an identity and a zero element with respect to functional composition $\circ$ clearly hold. Deflationarity holds by the semantics of PAL and BPAL. If $U_\varphi$ has an inverse element $U_\psi$, we must have $U_\varphi=U_\top$ since by deflationarity
\[M=U_\psi(U_\varphi(M))\preceq U_\varphi(M)\preceq M\] 
for all models $M$ and hence $U_\varphi=U_\top$. That is, one cannot cancel out a public announcement of $\varphi$ unless it is trivial. 

For commutativity, the formulas $p$ and $\Box_i p$ do not commute as can be checked for the model $M$ in Figure~\ref{fig:announcement-examples}. Neither left nor right absorption holds: $(U_p\circ U_q)(M)=(U_q\circ U_p)(M)\neq U_p(M)$. Monotonicity also fails since, for $N:=M_{|p}$, we have
\[
N\preceq M,
\qquad
U_{\Box_i p}(N)=N
\not\preceq
U_\bot(M)=U_{\Box_i p}(M).
\]

The formulas $\varphi:=p\land\Diamond_i\lnot p$ and $\psi:=(p\land\Diamond_i\lnot p)\land q$ commute: \(
U_\varphi\circ U_\psi(N)
=
U_\psi\circ U_\varphi(N)=U_\bot(N)
\) for all models $N$. In PAL, a public announcement of either $\varphi$ or $\psi$ leaves only $p$-states, making $\Diamond_i\lnot p$ false, so the next announcement eliminates all the states. However, $\varphi$ and $\psi$ do not strongly commute since a public announcement of $\varphi\land\psi$ leaves the single-state reflexive model $\{a\}$ (see the lower part of Figure~\ref{fig:announcement-examples}). Finally, idempotence does not hold, either: $U_\varphi^2(M)\neq U_\varphi(M)$.

\begin{figure}[htbp]
  \centering
  \begin{tikzpicture}[
    x=1cm, y=1cm,
    font=\small,
    line cap=round, line join=round,
    world/.style={circle, fill=black, inner sep=0pt,
                  minimum size=3.8pt},
    relation/.style={line width=0.65pt},
    announcement/.pic={
      \node[font=\Large, inner sep=0pt] {$\Longrightarrow$};
    }
  ]
    % Top panel: noncommutativity and failure of absorption.

    % Initial model.
    \node[world, label=below:$a$,
      label=above left:{$p,q$}] (a) at (0,-0.55) {};
    \node[world, label=below:$b$,
      label=above right:{$p,\neg q$}] (b) at (1.65,-0.55) {};
    \node[world, label=above left:$c$,
      label=above right:{$\neg p,\neg q$}]
      (c) at (0.825,0.87894) {};
    \draw[relation] (a) -- (b);
    \draw[relation] (b) -- (c);
    \draw[relation] (c) -- (a);
    \node at (0.825,-1.15) {$M$};

    % Initial announcements: p, Box_i p, and q.
    \begin{scope}[shift={(3.25,1.05)}, rotate=30]
      \pic[transform shape] {announcement};
    \end{scope}
    \node at (3.25,1.55) {$p$};

    \pic[transform shape] at (3.25,0) {announcement};
    \node at (3.25,0.35) {$\Box_i p$};

    \begin{scope}[shift={(3.25,-1.05)}, rotate=-30]
      \pic[transform shape] {announcement};
    \end{scope}
    \node at (3.25,-1.55) {$q$};

    % The model after announcing p.
    \node[world, label=below:$a$,
      label=above:{$p,q$}] (pa) at (4.60,1.45) {};
    \node[world, label=below:$b$,
      label=above:{$p,\neg q$}] (pb) at (6.25,1.45) {};
    \draw[relation] (pa) -- (pb);

    % Announce q after p.
    \begin{scope}[shift={(6.55,2.18)}, rotate=30]
      \pic[transform shape] {announcement};
    \end{scope}
    \node at (6.32,2.56) {$q$};
    \node[world, label=below:$a$,
      label=above:{$p,q$}] at (7.60,2.70) {};

    % Announce Box_i p after p.
    \pic[transform shape] at (7.40,1.45) {announcement};
    \node at (7.40,1.80) {$\Box_i p$};

    \node[world, label=below:$a$,
      label=above:{$p,q$}] (ppa) at (8.55,1.45) {};
    \node[world, label=below:$b$,
      label=above:{$p,\neg q$}] (ppb) at (10.20,1.45) {};
    \draw[relation] (ppa) -- (ppb);

    % Middle branch: announce Box_i p, then p.
    \node at (5.425,0) {$U_\bot(M)$};

    \pic[transform shape] at (7.40,0) {announcement};
    \node at (7.40,0.35) {$p$};
    \node at (9.375,0) {$U_\bot(M)$};

    % Lower branch: announce q, then p.
    \node[world, label=below:$a$,
      label=above:{$p,q$}] at (5.425,-1.45) {};

    \begin{scope}[shift={(6.55,-2.18)}, rotate=-30]
      \pic[transform shape] {announcement};
    \end{scope}
    \node at (6.32,-2.56) {$p$};
    \node[world, label=below:$a$,
      label=above:{$p,q$}] at (7.60,-2.70) {};

    % Bottom panel: commutativity without strong commutativity,
    % and failure of idempotence.
    \begin{scope}[yshift=-6.2cm]

      % The same initial model M.
      \node[world, label=below:$a$,
        label=above left:{$p,q$}] (a2) at (0,-0.55) {};
      \node[world, label=below:$b$,
        label=above right:{$p,\neg q$}] (b2) at (1.65,-0.55) {};
      \node[world, label=above left:$c$,
        label=above right:{$\neg p,\neg q$}]
        (c2) at (0.825,0.87894) {};
      \draw[relation] (a2) -- (b2);
      \draw[relation] (b2) -- (c2);
      \draw[relation] (c2) -- (a2);
      \node at (0.825,-1.15) {$M$};

      % First announce varphi.
      \begin{scope}[shift={(3.25,1.05)}, rotate=30]
        \pic[transform shape] {announcement};
      \end{scope}
      \node at (3.25,1.55) {$\varphi$};

      \node[world, label=below:$a$,
        label=above:{$p,q$}] (fa) at (4.60,1.45) {};
      \node[world, label=below:$b$,
        label=above:{$p,\neg q$}] (fb) at (6.25,1.45) {};
      \draw[relation] (fa) -- (fb);

      % Announce varphi again.
      \begin{scope}[shift={(7.40,1.85)}, rotate=30]
        \pic[transform shape] {announcement};
      \end{scope}
      \node at (7.25,2.30) {$\varphi$};
      \node at (9.375,2.15) {$U_\bot(M)$};

      % Announce psi after varphi.
      \begin{scope}[shift={(7.40,1.05)}, rotate=-30]
        \pic[transform shape] {announcement};
      \end{scope}
      \node at (7.25,0.60) {$\psi$};
      \node at (9.375,0.75) {$U_\bot(M)$};

      % Announce the conjunction directly.
      \node[font=\Large, inner sep=0pt] at (5.70,-0.55)
        {$\vphantom{\Longrightarrow}
          \smash{\xRightarrow{\hspace{5.30cm}}}$};
      \node at (5.70,-1.00) {$\varphi\land\psi$};

      \node[world, label=below:$a$,
        label=above:{$p,q$}] at (9.375,-0.55) {};
    \end{scope}
  \end{tikzpicture}

  \caption{Announcement updates in PAL, where
    \mbox{$\varphi=p\land\Diamond_i\neg p$} and
    \mbox{$\psi=\varphi\land q$}.
    Reflexive loops are omitted, and each edge represents
    accessibility in both directions for every agent.}
  \label{fig:announcement-examples}
\end{figure}
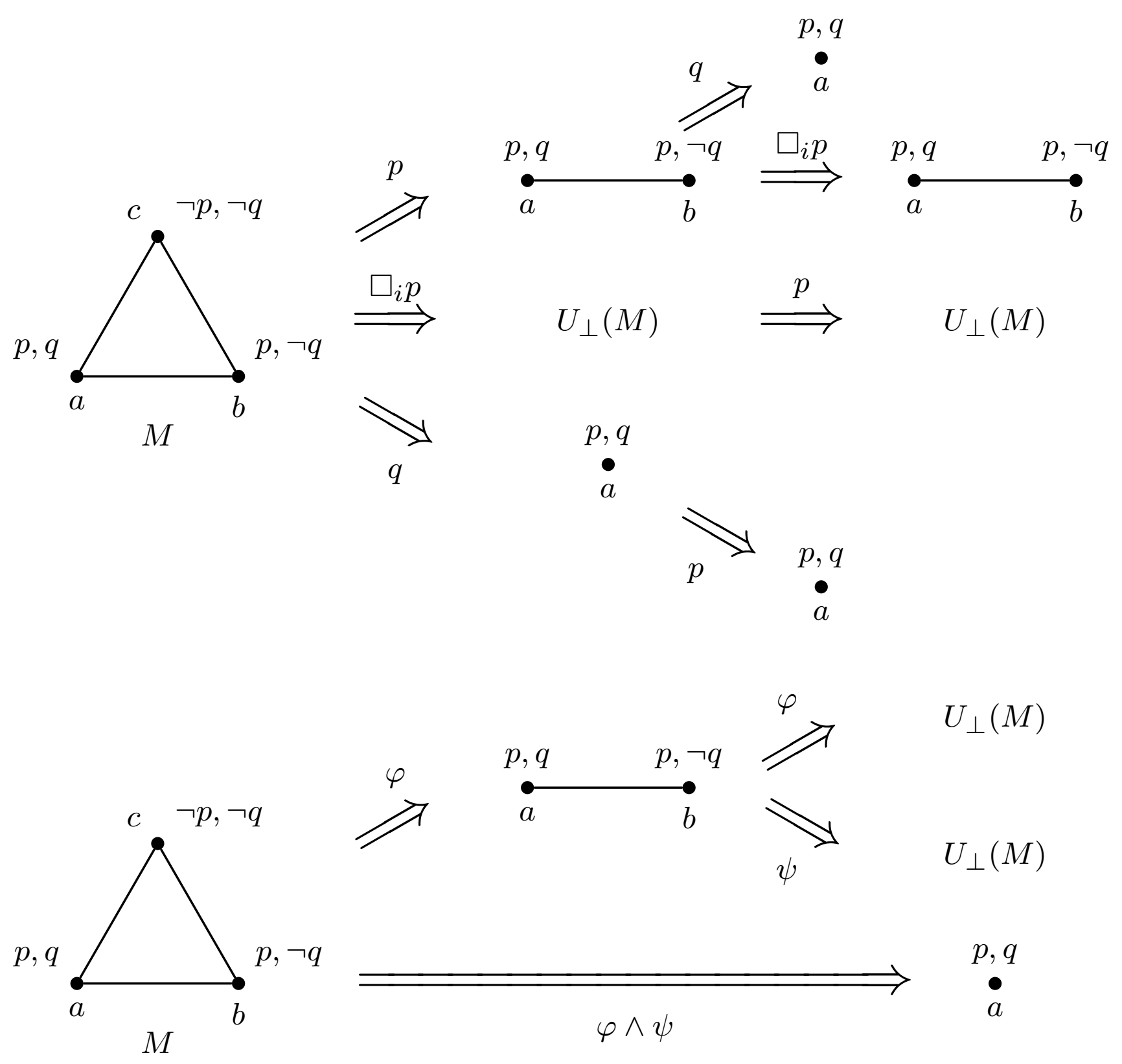
\begin{table}[htbp]
\centering
\caption{Algebraic properties of $U_\varphi$ in PAL and BPAL.}
\label{tab:algebraic-properties-announcement-operators}
\setlength{\tabcolsep}{3pt}
\begin{tabular}{p{0.20\textwidth} p{0.39\textwidth} p{0.13\textwidth} p{0.21\textwidth}}
\hline
\textbf{Property} & \textbf{Definition} & \raggedright\textbf{Whether it holds} & \raggedright\textbf{Whether it holds in the propositional case} \tabularnewline
\hline

Associativity
&
\(
(U_\varphi\circ U_\psi)\circ U_\chi
\)\newline
\(
=
U_\varphi\circ(U_\psi\circ U_\chi)
\)
&
Yes
&
Yes
\\\noalign{\vskip0.8em}

Identity
&
\(
U_\top\circ U_\varphi
=
U_\varphi\circ U_\top
=
U_\varphi
\)
&
Yes
&
Yes
\\\noalign{\vskip0.8em}

Zero element
&
\(
U_\bot\circ U_\varphi
=
U_\varphi\circ U_\bot
=
U_\bot
\)
&
Yes
&
Yes
\\\noalign{\vskip0.8em}

Inverse
&
\(
\forall U_\varphi\exists U_\psi\;
\)\newline
\(
\bigl(
U_\psi\circ U_\varphi
=
U_\varphi\circ U_\psi
=
U_\top
\bigr)
\)
&
No
&
No
\\\noalign{\vskip0.8em}

Commutativity
&
\(
U_\varphi\circ U_\psi
=
U_\psi\circ U_\varphi
\)
&
No
&
Yes
\\\noalign{\vskip0.8em}
Strong commutativity
&
\(
U_\varphi\circ U_\psi
=U_{\varphi\land \psi}=
U_\psi\circ U_\varphi
\)
&
No
&
Yes
\\\noalign{\vskip0.8em}
Idempotence
&
\(
U_\varphi\circ U_\varphi
=
U_\varphi
\)
&
No
&
Yes
\\\noalign{\vskip0.8em}

Left absorption
&
\(
U_\varphi\circ U_\psi
=
U_\varphi
\)
&
No
&
No
\\\noalign{\vskip0.8em}

Right absorption
&
\(
U_\psi\circ U_\varphi
=
U_\varphi
\)
&
No
&
No
\\\noalign{\vskip0.8em}

Monotonicity
&
\(
M\preceq N\Rightarrow
\)\newline
\(
U_\varphi(M)\preceq U_\varphi(N)
\)
&
No
&
Yes
\\\noalign{\vskip0.8em}

Deflationarity
&
\(
U_\varphi(M)\preceq M
\)
&
Yes
&
Yes
\\\noalign{\vskip0.8em}
\hline
\end{tabular}
\end{table}
\subsection{Commutativity, strong commutativity, and monotonicity}\label{subsec:commutativity-strong-commutativity-and-monotonicity}
In this subsection, we investigate the properties of commutativity, strong commutativity, and monotonicity.
\begin{lemma}\label{lem:sequential-announcements}
Let $\alpha,\beta$ be formulas. Both in PAL and BPAL,
\[U_\beta\circ U_\alpha=U_{\alpha\land[\alpha]\beta}.
\]
Moreover, for $\alpha,\beta\in\mathcal L$, the formulas $\alpha\land[!\alpha]\beta$ and $\alpha\land[\uparrow\alpha]\beta$ are logically equivalent in any class of frames.
\end{lemma}
\begin{proof}
Fix a model $M=(W,\{R_i\}_{i\in G},V)$. 

For PAL, a state $w\in W$ belongs to the domain of $(M_{|\alpha})_{|\beta}$ precisely when $M,w\models\alpha$ and $M_{|\alpha},w\models\beta$.
By the semantics of public announcement, this is equivalent to $M,w\models\alpha\land[!\alpha]\beta$.
Both successive restrictions and direct restriction to $\llbracket\alpha\land[!\alpha]\beta\rrbracket_M$ give the relations
\[
R_i\cap\bigl(
\llbracket\alpha\land[!\alpha]\beta\rrbracket_M
\times
\llbracket\alpha\land[!\alpha]\beta\rrbracket_M
\bigr)
\]
and the valuations $V(p)\cap\llbracket\alpha\land[!\alpha]\beta\rrbracket_M$.
Thus the two resulting models are equal, including when the retained domain is empty.

For BPAL, the relation for agent $i$ after both announcements is
\[
\begin{aligned}
&\bigl(R_i\cap(W\times\llbracket\alpha\rrbracket_M)\bigr)
\cap\bigl(W\times\llbracket\beta\rrbracket_{M|\alpha}\bigr)\\
&\qquad=R_i\cap\bigl(W\times(
\llbracket\alpha\rrbracket_M\cap\llbracket\beta\rrbracket_{M|\alpha}
)\bigr).
\end{aligned}
\]
Since the BPAL modality has no executability condition,
\[
\llbracket\beta\rrbracket_{M|\alpha}
=\llbracket[\uparrow\alpha]\beta\rrbracket_M.
\]
Consequently,
\[
\llbracket\alpha\rrbracket_M\cap\llbracket\beta\rrbracket_{M|\alpha}
=\llbracket\alpha\land[\uparrow\alpha]\beta\rrbracket_M.
\]
The domain and valuation remain unchanged in both constructions, proving the BPAL case.

For the final assertion, fix a model $M$.
At each $w\in\llbracket\alpha\rrbracket_M$, the models $M_{|\alpha}$ and $M|\alpha$ have the same atomic valuation and the same $i$-successors, namely the states $v\in\llbracket\alpha\rrbracket_M$ with $wR_iv$.
Induction on $\beta\in\mathcal L$ therefore gives $M_{|\alpha},w\models\beta$ if and only if $M|\alpha,w\models\beta$ for every $w\in\llbracket\alpha\rrbracket_M$.
The atomic and Boolean cases are immediate; in the modal case, all successors remain in $\llbracket\alpha\rrbracket_M$, so the induction hypothesis applies to every successor.
Thus the two guarded formulas agree at every $w\in\llbracket\alpha\rrbracket_M$.
At states outside $\llbracket\alpha\rrbracket_M$, both guarded formulas are false, proving their equivalence without any frame condition.
\end{proof}

\begin{lemma}[Announcement operators and validity]\label{lem:announcement-operators-and-validity}
For any formulas $\alpha,\beta$ and any class $\mathcal F$ of frames:
\begin{enumerate}
\item In PAL, $U_\alpha=U_\beta$ iff $\models_{\mathcal F}\alpha\leftrightarrow\beta$.
\item In BPAL, $U_\alpha=U_\beta$ iff $\models_{\mathcal F}\Box_i(\alpha\leftrightarrow\beta)$ for every $i\in G$.
\end{enumerate}
\end{lemma}
\begin{proof}
Fix a model $M=(W,\{R_i\}_{i\in G},V)$ in $\mathcal F$.

(1) Suppose $U_\alpha=U_\beta$. Then, $U_\alpha(M)=U_\beta(M)$, so $M_{|\alpha}=M_{|\beta}$, namely, $\llbracket\alpha\rrbracket_M=\llbracket\beta\rrbracket_M$ by the definition of updated models in PAL. Since $M$ was arbitrary, we have $\models_{\mathcal{F}}\alpha\leftrightarrow\beta$.
Conversely, suppose $\models_{\mathcal{F}}\alpha\leftrightarrow\beta$. Reversing the argument above yields $U_\alpha=U_\beta$.

(2) In BPAL, the domain and valuation are the same in both updates.
Their relations for agent $i$ are equal precisely when
\[
R_i\cap(W\times\llbracket\alpha\rrbracket_M)
=R_i\cap(W\times\llbracket\beta\rrbracket_M).
\]
This condition says that $v\in\llbracket\alpha\rrbracket_M$ if and only if $v\in\llbracket\beta\rrbracket_M$ for every $w,v\in W$ with $wR_iv$.
By the semantics of $\Box_i$, it is equivalent to $M,w\models\Box_i(\alpha\leftrightarrow\beta)$ for every $w\in W$.
Agreement at states that are not targets of any accessibility arrow is not required.
Taking all agents and all models based on frames in $\mathcal F$ proves both claims.
\end{proof}

\begin{lemma}[Equality of sequential announcement updates]\label{lem:equality-of-sequential-announcement-updates}
For a finite sequence $\sigma=(\varphi_1,\ldots,\varphi_n)$, write
$U_\sigma:=U_{\varphi_n}\circ\cdots\circ U_{\varphi_1}$,
$[!\sigma]\chi:=[!\varphi_1]\cdots[!\varphi_n]\chi$, and
$[\uparrow\sigma]\chi:=[\uparrow\varphi_1]\cdots[\uparrow\varphi_n]\chi$.
For any finite sequences $\sigma,\tau$ and any class $\mathcal F$ of frames:
\begin{enumerate}
\item In PAL, $U_\sigma=U_\tau$ iff
$\models_{\mathcal F}[!\sigma]\chi\leftrightarrow[!\tau]\chi$ for every $\chi\in\mathcal L_{PAL}$.
\item In BPAL, $U_\sigma=U_\tau$ iff
$\models_{\mathcal F}[\uparrow\sigma]\chi\leftrightarrow[\uparrow\tau]\chi$ for every $\chi\in\mathcal L_{\mathrm{BPAL}}$.
\end{enumerate}
\end{lemma}
\begin{proof}
(1)(2) The forward implications follow from the semantics. Conversely, suppose the relevant equivalence is valid for every $\chi$.
In PAL, taking $\chi=\bot$ shows that $U_\sigma(M)$ and $U_\tau(M)$ have the same domain for every model $M$ based on a frame in $\mathcal F$.
As induced submodels of $M$, they are therefore equal.

In BPAL, if $U_\sigma\ne U_\tau$, there is a model $M$ in $\mathcal{F}$ such that $U_\sigma(M)\neq U_\tau(M)$, namely $M|\sigma\neq M|\tau$. Then, there exist $i\in G$ and an edge $(x,y)\in R_i$ such that $M,y\models \sigma$ (defined as $M,y\models[\uparrow\varphi_1]\cdots[\uparrow\varphi_{k-1}]\varphi_k$ for every $1\leq k\leq n$, with the modality prefix empty when $k=1$; analogously for $\tau$) and $M,y\not\models\tau$, or $M,y\not\models \sigma$ and $M,y\models\tau$.
WLOG, assume the former and choose a proposition letter $q$ absent from $\sigma,\tau$ and set $V(q)=\{y\}$. Then $\chi=\Diamond_iq$ satisfies $M|\sigma,x\models\chi$ and $M|\tau,x\not\models\chi$, contradicting the assumed validity.
\end{proof}

\begin{theorem}[Commutativity]\label{thm:commutativity}
  The following are equivalent over any class of frames $\mathcal F$. In PAL:
  \begin{enumerate}
\item $U_\psi\circ U_\varphi=U_\varphi\circ U_\psi$
\item $\models_{\mathcal F}[!\psi][!\varphi]\chi\leftrightarrow[!\varphi][!\psi]\chi$ for every $\chi\in\mathcal L_{PAL}$.
\item 
$\models_{\mathcal F}
\bigl(
\varphi\land[!\varphi]\psi
\bigr)
\leftrightarrow
\bigl(
\psi\land[!\psi]\varphi
\bigr)$.
\item All of the following hold:
\begin{align*}
  \models_{\mathcal F}(\varphi\land\neg\psi)
   \rightarrow[!\varphi]\neg\psi\\\models_{\mathcal F}(\psi\land\neg\varphi)
   \rightarrow[!\psi]\neg\varphi\\
   \models_{\mathcal F}(\varphi\land\psi)
   \rightarrow
   \bigl([!\varphi]\psi
          \leftrightarrow
          [!\psi]\varphi\bigr).
\end{align*}
  \end{enumerate}
  In BPAL:
  \begin{enumerate}
\item $U_\psi\circ U_\varphi=U_\varphi\circ U_\psi$
\item $\models_{\mathcal F}[\uparrow\varphi][\uparrow\psi]\chi\leftrightarrow[\uparrow\psi][\uparrow\varphi]\chi$ for every $\chi\in\mathcal L_{\mathrm{BPAL}}$.
\item 
$\models_{\mathcal F}\Box_i(
\bigl(
\varphi\land[\uparrow\varphi]\psi
\bigr)
\leftrightarrow
\bigl(
\psi\land[\uparrow\psi]\varphi
\bigr))$ for all $i\in G$.
\item All of the following hold for all $i\in G$:
\begin{align*}
  \models_{\mathcal F}\Box_i((\varphi\land\neg\psi)
   \rightarrow[\uparrow\varphi]\neg\psi)\\\models_{\mathcal F}\Box_i((\psi\land\neg\varphi)
   \rightarrow[\uparrow\psi]\neg\varphi)\\
   \models_{\mathcal F}\Box_i((\varphi\land\psi)
   \rightarrow
   \bigl([\uparrow\varphi]\psi
          \leftrightarrow
          [\uparrow\psi]\varphi\bigr)).
\end{align*}
  \end{enumerate}

If $\mathcal F$ is reflexive (e.g., S5), the outer $\Box_i$ in the BPAL conditions may be omitted.
If, in addition, $\varphi,\psi\in\mathcal L$, the PAL conditions and the BPAL conditions above are equivalent to one another.
\end{theorem}
\begin{proof}
\emph{PAL.} Let $\sigma=(\varphi,\psi)$ and $\tau=(\psi,\varphi)$.
By Lemma~\ref{lem:equality-of-sequential-announcement-updates}, conditions (1) and (2) are equivalent.
Lemma~\ref{lem:sequential-announcements} gives $U_\psi\circ U_\varphi=U_{\varphi\land[!\varphi]\psi}$ and $U_\varphi\circ U_\psi=U_{\psi\land[!\psi]\varphi}$.
By the PAL part of Lemma~\ref{lem:announcement-operators-and-validity}, condition (1) is equivalent to
$\models_{\mathcal F}\varphi\land[!\varphi]\psi\leftrightarrow\psi\land[!\psi]\varphi$, which is condition (3).

For the equivalence of (3) and (4), fix a model $M$ based on a frame in $\mathcal F$ and a state $w$, and consider the four possible initial truth-value combinations of $\varphi$ and $\psi$.
If $M,w\models\varphi\land\neg\psi$, then $\psi\land[!\psi]\varphi$ is false at $w$, whereas $\varphi\land[!\varphi]\psi$ has the truth value of $[!\varphi]\psi$.
Since $\varphi$ is true at $w$, agreement of $\varphi\land[!\varphi]\psi$ and $\psi\land[!\psi]\varphi$ is therefore equivalent to $M,w\models[!\varphi]\neg\psi$.
This is exactly the requirement in the first implication of (4) when its antecedent holds.
If $M,w\models\psi\land\neg\varphi$, the same argument with $\varphi$ and $\psi$ interchanged gives the second implication.
If both formulas are true at $w$, agreement of $\varphi\land[!\varphi]\psi$ and $\psi\land[!\psi]\varphi$ is precisely
$M,w\models[!\varphi]\psi\leftrightarrow[!\psi]\varphi$, as required by the third implication.
Finally, if both formulas are false, then both $\varphi\land[!\varphi]\psi$ and $\psi\land[!\psi]\varphi$ are false, and all three implications are vacuously true.
These cases exhaust all possibilities, so condition (3) is equivalent to condition (4).

\emph{BPAL.} Again let $\sigma=(\varphi,\psi)$ and $\tau=(\psi,\varphi)$.
By Lemma~\ref{lem:equality-of-sequential-announcement-updates}, conditions (1) and (2) are equivalent.
The BPAL parts of Lemmas~\ref{lem:sequential-announcements} and~\ref{lem:announcement-operators-and-validity} identify condition (1) with
$\models_{\mathcal F}\Box_i((\varphi\land[\uparrow\varphi]\psi)\leftrightarrow(\psi\land[\uparrow\psi]\varphi))$ for every $i\in G$, which is condition (3).
To obtain (4), apply the same four-case analysis at an arbitrary state, now using the BPAL modality.
Here $[\uparrow\alpha]\neg\beta$ is equivalent to $\neg[\uparrow\alpha]\beta$, since the updated model is evaluated without an executability condition.
Consequently, $(\varphi\land[\uparrow\varphi]\psi)\leftrightarrow(\psi\land[\uparrow\psi]\varphi)$ is pointwise equivalent to the conjunction of the three implications inside the boxes in (4).
For each $i$, a conjunction holds at every $i$-successor exactly when each conjunct does.
Applying this observation at every state of every model based on a frame in $\mathcal F$ proves the equivalence of (3) and (4) in BPAL.

\emph{Reflexive initial frames.} Suppose that every frame in $\mathcal F$ is reflexive.
For any formula $\delta$, $\models_{\mathcal F}\delta$ implies $\models_{\mathcal F}\Box_i\delta$ for every $i$, by the semantics of $\Box_i$.
Conversely, fix $i\in G$; if $\Box_i\delta$ holds at every state of every model based on a frame in $\mathcal F$, then $\delta$ does as well, because $wR_iw$.
This proves that all outer boxes in the BPAL conditions may be omitted.
Now assume also $\varphi,\psi\in\mathcal L$.
The last assertion of Lemma~\ref{lem:sequential-announcements} identifies $\varphi\land[!\varphi]\psi$ with $\varphi\land[\uparrow\varphi]\psi$, and $\psi\land[!\psi]\varphi$ with $\psi\land[\uparrow\psi]\varphi$, at every state of every initial model.
Hence PAL condition (3) is equivalent to BPAL condition (3) after removal of its outer boxes.
Together with the equivalences already proved within each logic, this establishes the equivalence of the PAL and BPAL conditions, in particular over S5 frames.
\end{proof}
Note that the commutativity of public announcement operators (1) coincides with the formula-level commutativity (2).
Also, condition (3) on the PAL side can be written using the dual as $\models\langle !\varphi\rangle\psi\leftrightarrow\langle !\psi\rangle\varphi$. This cannot be done on the BPAL side.
\begin{theorem}[Strong commutativity]\label{thm:strong-commutativity}The following are equivalent over any class of frames $\mathcal F$. In PAL:
  \begin{enumerate}
  \item \[
U_\psi\circ U_\varphi
=
U_{\varphi\land\psi}
=
U_\varphi\circ U_\psi
\]
\item \[
\models_{\mathcal F}
[!\varphi][!\psi]\chi
\leftrightarrow
[!(\varphi\land\psi)]\chi
\text{ and }
\models_{\mathcal F}
[!(\varphi\land\psi)]\chi
\leftrightarrow
[!\psi][!\varphi]\chi
\]
for every $\chi\in \mathcal{L}_{PAL}$
\item \[
\models_{\mathcal F}
\varphi\rightarrow
\bigl(\psi\leftrightarrow[!\varphi]\psi\bigr)
\]
and
\[
\models_{\mathcal F}
\psi\rightarrow
\bigl(\varphi\leftrightarrow[!\psi]\varphi\bigr)
\]
  \end{enumerate}
  In BPAL:
  \begin{enumerate}
  \item \[
U_\psi\circ U_\varphi
=
U_{\varphi\land\psi}
=
U_\varphi\circ U_\psi
\]
\item \[
\models_{\mathcal F}
[\uparrow\varphi][\uparrow\psi]\chi
\leftrightarrow
[\uparrow(\varphi\land\psi)]\chi
\text{ and }
\models_{\mathcal F}
[\uparrow(\varphi\land\psi)]\chi
\leftrightarrow
[\uparrow\psi][\uparrow\varphi]\chi
\]
for every $\chi\in \mathcal{L}_{BPAL}$

\item For all $i\in G$, \[
\models_{\mathcal F}
\Box_i(\varphi\rightarrow
\bigl(\psi\leftrightarrow[\uparrow\varphi]\psi\bigr))
\]
\[
\models_{\mathcal F}\Box_i(
\psi\rightarrow
\bigl(\varphi\leftrightarrow[\uparrow\psi]\varphi\bigr))
\]
  \end{enumerate}

If every frame in $\mathcal F$ is reflexive (e.g., S5), the outer $\Box_i$ in the BPAL conditions may be omitted.
If, in addition, $\varphi,\psi\in\mathcal L$, the PAL conditions and the BPAL conditions above are equivalent to one another.
\end{theorem}
\begin{proof}
\emph{PAL.} Condition (1) is equivalent to the conjunction of
$U_\psi\circ U_\varphi=U_{\varphi\land\psi}$ and
$U_\varphi\circ U_\psi=U_{\varphi\land\psi}$.
Apply Lemma~\ref{lem:equality-of-sequential-announcement-updates} first to
$\sigma=(\varphi,\psi)$ and $\tau=(\varphi\land\psi)$, and then to
$\sigma=(\psi,\varphi)$ and $\tau=(\varphi\land\psi)$.
The two operator equalities are respectively equivalent to the two language-level equivalences in condition (2).
Hence (1) and (2) are equivalent.

By Lemmas~\ref{lem:sequential-announcements} and~\ref{lem:announcement-operators-and-validity},
the equality $U_\psi\circ U_\varphi=U_{\varphi\land\psi}$ is equivalent to the validity of
$(\varphi\land[!\varphi]\psi)\leftrightarrow(\varphi\land\psi)$ on $\mathcal F$.
By propositional reasoning, this is equivalent to
$\varphi\rightarrow(\psi\leftrightarrow[!\varphi]\psi)$, which is the first formula in condition (3).
Interchanging $\varphi$ and $\psi$ shows that
$U_\varphi\circ U_\psi=U_{\varphi\land\psi}$ is equivalent to the second formula in condition (3);
here $\psi\land\varphi$ and $\varphi\land\psi$ define the same restriction.
Thus (1) and (3) are equivalent.

\emph{BPAL.} As in PAL, condition (1) is the conjunction of
$U_\psi\circ U_\varphi=U_{\varphi\land\psi}$ and
$U_\varphi\circ U_\psi=U_{\varphi\land\psi}$.
Applying the BPAL part of Lemma~\ref{lem:equality-of-sequential-announcement-updates} to the same pairs of sequences shows that these two equalities are respectively equivalent to the two language-level equivalences in condition (2).
Hence (1) and (2) are equivalent.

The BPAL parts of Lemmas~\ref{lem:sequential-announcements} and~\ref{lem:announcement-operators-and-validity} show that
$U_\psi\circ U_\varphi=U_{\varphi\land\psi}$ is equivalent to
$\models_{\mathcal F}\Box_i((\varphi\land[\uparrow\varphi]\psi)\leftrightarrow(\varphi\land\psi))$
for every $i\in G$.
The formula inside the box is propositionally equivalent to
$\varphi\rightarrow(\psi\leftrightarrow[\uparrow\varphi]\psi)$,
so this is exactly the first family of validities in BPAL condition (3).
Interchanging $\varphi$ and $\psi$ gives the second family, corresponding to
$U_\varphi\circ U_\psi=U_{\varphi\land\psi}$.
Thus (1) and (3) are equivalent in BPAL as well.

\emph{Reflexive initial frames.} The argument in Theorem~\ref{thm:commutativity} allows the outer boxes in condition (3) to be omitted when every frame in $\mathcal F$ is reflexive.
If also $\varphi,\psi\in\mathcal L$, Lemma~\ref{lem:sequential-announcements} shows that
$\varphi\land[!\varphi]\psi$ and $\varphi\land[\uparrow\varphi]\psi$ agree everywhere.
Therefore their respective biconditionals with $\varphi\land\psi$ also agree everywhere.
By the propositional equivalence used above, this identifies the first PAL stability condition with the first unboxed BPAL stability condition.
Interchanging $\varphi$ and $\psi$ does the same for the second conditions.
This proves the equivalence of the PAL and BPAL assertions over reflexive frame classes, including S5 frames.
\end{proof}
The following corollary reveals the gap between commutativity and strong commutativity.
\begin{corollary}[Commutativity and strong commutativity]\label{cor:commutativity-and-strong-commutativity}
Let $\mathcal F$ be any class of frames.
\begin{enumerate}
\item In PAL, $U_\varphi$ and $U_\psi$ strongly commute iff they commute and
\[
\models_{\mathcal F}(\varphi\land\psi)\to
\bigl([!\varphi]\psi\land[!\psi]\varphi\bigr).
\]
\item In BPAL, $U_\varphi$ and $U_\psi$ strongly commute iff they commute and, for every $i\in G$,
\[
\models_{\mathcal F}\Box_i\bigl((\varphi\land\psi)\to
\bigl([\uparrow\varphi]\psi\land[\uparrow\psi]\varphi\bigr)\bigr).
\]
\end{enumerate}
\end{corollary}
\begin{proof}
(1)
$(\Rightarrow)$
Suppose that $U_\varphi$ and $U_\psi$ strongly commute.
They commute by definition.
By Theorem~\ref{thm:strong-commutativity},
$\models_{\mathcal F}(\varphi\land[\varphi]\psi)\leftrightarrow(\varphi\land\psi)$, $\models_{\mathcal F}(\psi\land[\psi]\varphi)\leftrightarrow(\varphi\land\psi)$.
In particular,
\[
\models_{\mathcal F}(\varphi\land\psi)\to((\varphi\land[\varphi]\psi)\land(\psi\land[\psi]\varphi)).
\]
This is
equivalent to the mutual-preservation condition
\[
\models_{\mathcal F}(\varphi\land\psi)\to
\bigl([!\varphi]\psi\land[!\psi]\varphi\bigr).
\]
$(\Leftarrow)$
Suppose that $U_\varphi$ and $U_\psi$ commute and that the
mutual-preservation condition holds.
By Theorem~\ref{thm:commutativity},
$\models_{\mathcal F}(\varphi\land[\varphi]\psi)\leftrightarrow(\psi\land[\psi]\varphi)$.
Thus, by propositional reasoning
we have
$\models_{\mathcal F}(\varphi\land[\varphi]\psi)\to(\varphi\land\psi)$ and
$\models_{\mathcal F}(\psi\land[\psi]\varphi)\to(\varphi\land\psi)$.
The mutual-preservation assumption gives the reverse implications,
since it is equivalent to
\[
\models_{\mathcal F}(\varphi\land\psi)\to((\varphi\land[\varphi]\psi)\land(\psi\land[\psi]\varphi)).
\]
Consequently,
\[
\models_{\mathcal F}(\varphi\land[\varphi]\psi)\leftrightarrow(\varphi\land\psi),
\qquad
\models_{\mathcal F}(\psi\land[\psi]\varphi)\leftrightarrow(\varphi\land\psi).
\]
Theorem~\ref{thm:strong-commutativity} now yields strong
commutativity.

(2)
$(\Rightarrow)$
Suppose that $U_\varphi$ and $U_\psi$ strongly commute.
They commute by definition.
By Theorem~\ref{thm:strong-commutativity},
$(\varphi\land[\varphi]\psi)\leftrightarrow(\varphi\land\psi)$ and $(\psi\land[\psi]\varphi)\leftrightarrow(\varphi\land\psi)$
hold at every $i$-successor of every state in every
$\mathcal F$-model, for each $i\in G$.
Hence $(\varphi\land\psi)\to((\varphi\land[\varphi]\psi)\land(\psi\land[\psi]\varphi))$ holds at every such successor.
This is precisely the boxed mutual-preservation condition.

\smallskip
\noindent
$(\Leftarrow)$
Suppose that $U_\varphi$ and $U_\psi$ commute and that the
boxed mutual-preservation condition holds.
By Theorem~\ref{thm:commutativity} and the preservation assumption,
for every $i\in G$,
\[
\models_{\mathcal F}\Box_i((\varphi\land[\varphi]\psi)\leftrightarrow(\psi\land[\psi]\varphi)),
\qquad
\models_{\mathcal F}\Box_i
\bigl((\varphi\land\psi)\to((\varphi\land[\varphi]\psi)\land(\psi\land[\psi]\varphi))\bigr).
\]
Applying the same propositional reasoning as in the PAL case
at each $i$-successor yields
\[
\models_{\mathcal F}\Box_i
\bigl(((\varphi\land[\varphi]\psi)\leftrightarrow(\varphi\land\psi))
      \land((\psi\land[\psi]\varphi)\leftrightarrow(\varphi\land\psi))\bigr)
\qquad(i\in G).
\]
Theorem~\ref{thm:strong-commutativity}
therefore yields strong commutativity.
\end{proof}
We next define several monotonicity notions. Positive and negative monotonicity are due to \citet{masuzawaHasebe2011}. 
\begin{definition}\label{def:monotonicity-in-announcements}
  Let $\varphi$ be a formula.
  \begin{itemize}
    \item $\varphi$ is \emph{monotone} (\emph{in announcements}) iff $\forall\alpha,\beta
(\models\alpha\to\beta$ implies $\models[\alpha]\varphi\to[\beta]\varphi$)
    \item $\varphi$ is \emph{positively monotone} iff $\forall\alpha,\beta
(\models\alpha\to\beta$ implies $\models [\beta]\varphi\to[\alpha]\varphi$)
\item $\varphi$ is \emph{negatively monotone} iff $
\forall\alpha,\beta(\models\alpha\to\beta$ implies $\models\langle \alpha\rangle\varphi\to\langle\beta\rangle\varphi$)
  \end{itemize}
\end{definition}
Note that monotonicity in the postcondition $\forall\alpha,\beta
(\models\alpha\to\beta$ implies $\models[\varphi]\alpha\to[\varphi]\beta$) always holds over K both in PAL and BPAL by necessitation ($\models_{\mathrm{K}} \psi$ implies $\models_{\mathrm{K}} [\varphi]\psi$ for all $\varphi$) and distribution $\models[\varphi](\alpha\to\beta)\to
\bigl([\varphi]\alpha\to[\varphi]\beta\bigr)$ of the public announcement modality $[\cdot]$.

We next define deletion monotonicity and deletion antimonotonicity, which correspond to downward and upward monotonicity in the accessibility relations (\citealp[Definition~4.5]{kontinenVaananen2009} and \citealp[Definition~7.4.1]{deRijke1993}).
\begin{definition}\label{def:deletion-monotonicity}
    Let $\varphi$ be a formula.
    \begin{itemize}
        \item $\varphi$ is \textit{deletion-monotone} iff for all models $M=(W,\{R_i\}_{i\in G},V)$, $M'=(W, \{R_i'\}_{i\in G}, V)$ with $R_i'\subseteq R_i$ for all $i\in G$, and states $w\in W$, $M,w\models \varphi$ implies $M',w\models \varphi$.
        \item $\varphi$ is \textit{deletion-antimonotone} iff for all models $M=(W,\{R_i\}_{i\in G},V)$, $M'=(W, \{R_i'\}_{i\in G}, V)$ with $R_i'\subseteq R_i$ for all $i\in G$, and states $w\in W$, $M',w\models \varphi$ implies $M,w\models \varphi$.
    \end{itemize} 
\end{definition}
We recall the notions of bisimulation, simulation, and refinement. For a comprehensive study of simulations and refinements, see~\citet{bozzelliEtAl2014}.
\begin{samepage}
\begin{definition}[Bisimulation, simulation, and refinement]\leavevmode\label{def:bisimulation-simulation-and-refinement}
Let
$M=(W,\{R_i\}_{i\in G},V)$ and $N=(W',\{R_i'\}_{i\in G},V')$
be models, and let $Z\subseteq W\times W'$ be a nonempty relation.
Consider the following conditions, required whenever $xZy$:
\begin{itemize}
\item \emph{Atoms:} for every $q\in\mathbf{Prop}$, $x\in V(q)$ iff $y\in V'(q)$.
\item \emph{Forth:} for every $i\in G$ and $x'$ with $xR_ix'$, there is a $y'$ with $yR_i'y'$ and $x'Zy'$.
\item \emph{Back:} for every $i\in G$ and $y'$ with $yR_i'y'$, there is an $x'$ with $xR_ix'$ and $x'Zy'$.
\end{itemize}
The relation $Z$ is a \emph{bisimulation} from $M$ to $N$ if it satisfies all three conditions, a \emph{simulation} if it satisfies Atoms and Forth, and a \emph{refinement} if it satisfies Atoms and Back.
The pointed models $(M,w)$ and $(N,v)$ are \emph{bisimilar} if a bisimulation contains $(w,v)$.
The pointed model $(N,v)$ is a \emph{simulation} (respectively, a \emph{refinement}) of $(M,w)$ if a simulation (respectively, a refinement) from $M$ to $N$ contains $(w,v)$.
A formula $\varphi$ is \emph{preserved under simulations} (respectively, \emph{refinements}) if $M,w\models\varphi$ implies $N,v\models\varphi$ whenever $(N,v)$ is a simulation (respectively, a refinement) of $(M,w)$.
\end{definition}
\end{samepage}
Note that $Z$ is a simulation from $M,w$ to $N,v$ iff $Z^{-1}$ is a refinement from $N,v$ to $M,w$.
\begin{definition}[\citealp{hollidayIcard2010}]\label{def:universal-and-existential-formulas}
  Define \emph{universal formulas} and \emph{existential formulas}, respectively, by
  \[
  \varphi::=p\mid \lnot p\mid \varphi\land\psi\mid \varphi\lor\psi\mid \Box_i\varphi\quad (i\in G, \,p\in\mathbf{Prop})
  \]
    \[
  \varphi::=p\mid \lnot p\mid \varphi\land\psi\mid \varphi\lor\psi\mid \Diamond_i\varphi \quad (i\in G,\, p\in\mathbf{Prop}).
  \]
\end{definition}
The following theorem combines standard preservation results with
characterizations in terms of announcement operators.
In each list, $(2)\Leftrightarrow(4)$ is the classical modal
preservation theorem recalled in
\citet[Theorem~2.4 and the subsequent discussion]{hollidayIcard2010}.
The equivalences with (1) and (3) follow from this theorem and
elementary preservation arguments; see also the treatment of
refinements and simulations in \citet{bozzelliEtAl2014}.
Positive and negative monotonicity, together with syntactic
sufficient conditions, were studied in an extension of PAL
by \citet{masuzawaHasebe2011}.
For PAL, the implication $(4)\Rightarrow(6)$ in the second list
is established in \citet{vanBenthem2007}, and its first-list
counterpart follows by applying this result to $\neg\varphi$.
Our contribution here is a unified characterization over
K for both PAL and BPAL: we establish the converse
implications $(5)\Rightarrow(4)$ and $(6)\Rightarrow(4)$,
and connect these properties with the operator comparisons
in (7) and the announcement schemata in (8). 
\begin{theorem}[Monotonicity]\label{thm:monotonicity}
    Let $\varphi$ be a formula of the relevant announcement language. Then the following are equivalent over K, both in PAL and BPAL:
    \begin{enumerate}
        \item $\varphi$ is deletion-monotone.
        \item $\varphi$ is preserved under submodels.
        \item $\varphi$ is preserved under refinements.
        \item $\varphi\equiv \psi$ for some universal formula $\psi$.
        \item $\varphi$ is positively monotone.
        \item $U_{\lnot\varphi}$ is monotone.
        \item $U_{\varphi\land\psi}(M)\preceq (U_\varphi\circ U_\psi)(M)$ for all formulas $\psi$ and models $M$.
        \item $\models\varphi\to[\psi]\varphi$ for every formula $\psi$.
    \end{enumerate}
    Furthermore, the following are equivalent over K, both in PAL and BPAL:
    \begin{enumerate}
        \item $\varphi$ is deletion-antimonotone.
        \item $\varphi$ is preserved under model extensions.
        \item $\varphi$ is preserved under simulations.
        \item $\varphi\equiv \psi$ for some existential formula $\psi$.
        \item $\varphi$ is negatively monotone.
        \item $U_\varphi$ is monotone. 
        \item $(U_\varphi\circ U_\psi)(M)\preceq U_{\varphi\land\psi}(M)$ for all formulas $\psi$ and models $M$.
        \item $\models\langle\psi\rangle\varphi\to\varphi$ for every formula $\psi$.
    \end{enumerate}
    Finally, in PAL: 
    \[\varphi\text{ is monotone }\Longleftrightarrow \varphi\text{ is valid}
    \]
    and in BPAL:
    \[\varphi \text{ is monotone }\Longleftrightarrow \varphi\text{ is negatively monotone}.
    \]
\end{theorem}
\begin{proof}
Reduction axioms replace $\varphi$ by an equivalent formula in $\mathcal L$ without introducing proposition letters, so we may assume $\varphi\in\mathcal L$.
All the stated properties are invariant under this replacement.
For the converse implications below, fix a proposition letter $p$ not occurring in $\varphi$; its valuation can specify an arbitrary set of states without changing the truth of $\varphi$.
We first prove the equivalences in the first list.

\emph{$(1)\Rightarrow(2)\Rightarrow(4)\Rightarrow(3)\Rightarrow(1)$.}
For an induced submodel $N=M\upharpoonright S$ and $w\in S$, put
\[
M^\sharp=(W,\{R_i\cap(S\times S)\}_{i\in G},V).
\]
The models $M^\sharp$ and $N$ agree on the generated submodel at $w$.
Thus deletion-monotonicity transfers truth from $M,w$ to $M^\sharp,w$, and then to $N,w$, proving $(1)\Rightarrow(2)$.
The implication $(2)\Rightarrow(4)$ is the modal submodel-preservation theorem of \citet[Theorem~2.4]{hollidayIcard2010}.
Universal formulas are preserved under refinements by structural induction: literals use atoms, the Boolean cases are immediate, and the box case uses back and the induction hypothesis.
This proves $(4)\Rightarrow(3)$.
Finally, the identity relation witnesses that every same-domain arrow deletion is a refinement, giving $(3)\Rightarrow(1)$.

\emph{$(4)\Rightarrow(5)\Rightarrow(8)\Rightarrow(2)$.}
If $\models\alpha\to\beta$, then $U_\alpha(M)\preceq U_\beta(M)$.
Preservation of universal formulas therefore gives $\models[\beta]\varphi\to[\alpha]\varphi$; in PAL, at states not satisfying $\alpha$, the consequent is vacuously true.
Applying positive monotonicity to $\psi\to\top$ and using $[\top]\varphi\equiv\varphi$ yields (8).
Conversely, assume (8), let $N=M\upharpoonright S$ and $w\in S$, and interpret the fresh letter $p$ by $S$.
This does not affect $\varphi$ in either $M$ or $N$.
In PAL, $U_p(M)=N$; in BPAL, $U_p(M),w$ and $N,w$ have the same generated submodel.
Applying (8) with $\psi=p$, truth of $\varphi$ at $M,w$ therefore implies truth at $N,w$, proving (2).

\emph{Condition (6).}
In PAL, let $N\preceq M$.
Since both updated models are induced submodels of $M$, $U_{\neg\varphi}(N)\preceq U_{\neg\varphi}(M)$ holds exactly when every state retained in $U_{\neg\varphi}(N)$ is also retained in $U_{\neg\varphi}(M)$.
Thus monotonicity of $U_{\neg\varphi}$ is equivalent to requiring that $N,w\models\neg\varphi$ implies $M,w\models\neg\varphi$ for every $N\preceq M$ and every state $w$ of $N$.
By contraposition, this is equivalent to requiring that $M,w\models\varphi$ implies $N,w\models\varphi$ for all such $N,M,w$, which is precisely preservation under submodels, i.e., (2).
In BPAL, let $N\preceq_B M$. The model $N$ can be obtained by first taking the induced restriction of $M$ to the domain of $N$ and then deleting arrows.
Conditions (1) and (2) therefore imply that $M,w\models\varphi$ entails $N,w\models\varphi$ for every state $w$ of $N$.
By contraposition, every arrow retained by $U_{\neg\varphi}(N)$ is also retained by $U_{\neg\varphi}(M)$, proving monotonicity with respect to $\preceq_B$.
For the converse, let $N\preceq_B M$ have the same domain and valuation, and suppose $N,w\models\neg\varphi$.
Fix $i\in G$ and add to both models a new state $z$ and only the new arrow $zR_iw$, obtaining $N^+\preceq_B M^+$.
Use identical atomic valuations at $z$ in the two models.
No old state can reach $z$, so truth at old states is unchanged.
The arrow $(z,w)$ survives $U_{\neg\varphi}(N^+)$; monotonicity forces it to survive $U_{\neg\varphi}(M^+)$ as well.
Thus $M,w\models\neg\varphi$, which, by contraposition, proves (1).

\emph{$(4)\Rightarrow(7)\Rightarrow(2)$.}
Fix a formula $\psi$ and a model $M$.
The updates $U_{\varphi\land\psi}(M)$ and $U_\varphi(U_\psi(M))$ retain, respectively, $\llbracket\psi\rrbracket_M\cap\llbracket\varphi\rrbracket_M$ and $\llbracket\psi\rrbracket_M\cap\llbracket\varphi\rrbracket_{U_\psi(M)}$ as their domains in PAL and as their permitted arrow targets in BPAL.
Universal preservation gives $\llbracket\psi\rrbracket_M\cap\llbracket\varphi\rrbracket_M\subseteq\llbracket\psi\rrbracket_M\cap\llbracket\varphi\rrbracket_{U_\psi(M)}$, proving (7).
Conversely, suppose $N$ is an induced submodel of $M$, $w$ is a state of $N$, and $M,w\models\varphi$, and interpret $p$ by the domain of $N$.
Apply (7) with $\psi=p$.
In PAL, $w$ belongs to $U_{\varphi\land p}(M)$, so it belongs to $U_\varphi(U_p(M))$, and hence $N,w\models\varphi$.
In BPAL, add a fresh predecessor $zR_iw$ as above, with $p$ false at $z$, and call the resulting model $M^+$.
The arrow $(z,w)$ survives $U_{\varphi\land p}(M^+)$, so (7) forces it to survive $U_\varphi(U_p(M^+))$.
Hence $U_p(M^+),w\models\varphi$.
The generated submodel at $w$ in $U_p(M^+)$ is the same as in $N$, so $N,w\models\varphi$, proving (2).

\emph{The second list.}
Apply the first list to $\neg\varphi$.
Submodel preservation dualizes to preservation under model extensions, deletion-monotonicity to deletion-antimonotonicity, and universal definability to existential definability.
For condition (3), $Z$ is a simulation from $M$ to $N$ containing $(w,v)$ iff $Z^{-1}$ is a refinement from $N$ to $M$ containing $(v,w)$.
Thus, by contraposition, preservation of $\neg\varphi$ under refinements is exactly preservation of $\varphi$ under simulations.
By contraposition and announcement duality, positive monotonicity of $\neg\varphi$ is exactly negative monotonicity of $\varphi$.
The operator-monotonicity condition becomes monotonicity of $U_\varphi$.
For condition (7), fix arbitrary $\psi$ and $M$.
The two retention sets in the universal comparison for $\neg\varphi$ are $\llbracket\psi\rrbracket_M\setminus\llbracket\varphi\rrbracket_M$ and $\llbracket\psi\rrbracket_M\setminus\llbracket\varphi\rrbracket_{U_\psi(M)}$.
Their inclusion is equivalent to $\llbracket\psi\rrbracket_M\cap\llbracket\varphi\rrbracket_{U_\psi(M)}\subseteq\llbracket\psi\rrbracket_M\cap\llbracket\varphi\rrbracket_M$ in PAL.
In BPAL, the same complementation argument applies separately at the target of every original arrow.
Thus the universal comparison for $\neg\varphi$ is equivalent to
\[
(U_\varphi\circ U_\psi)(M)\preceq U_{\varphi\land\psi}(M),
\]
as required.
For condition (8), contraposition and announcement duality give
\[
\models\neg\varphi\to[\psi]\neg\varphi
\quad\Longleftrightarrow\quad
\models\langle\psi\rangle\varphi\to\varphi
\]
for every formula $\psi$.

Finally, in PAL, monotonicity in announcements applied to $\bot\to\top$ yields $\models[!\bot]\varphi\to[!\top]\varphi$, and hence $\models\varphi$.
The converse follows because a valid formula holds after every executable announcement.
In BPAL, $[\alpha]\varphi$ and $\langle\alpha\rangle\varphi$ are equivalent, so monotonicity and negative monotonicity in announcements coincide.
\end{proof}
Refinement is more general than model restriction, allowing states
to be duplicated as well as eliminated.
More precisely, up to bisimulation, a refinement can be obtained
by first passing to a bisimilar model and then taking an induced
submodel \citep[Section~3.3]{bozzelliEtAl2014}.
Since PAL and BPAL formulas are invariant under bisimulation,
preservation under submodels already ensures preservation under
refinements; the converse holds because every induced submodel
is a refinement.
Thus, over K, these operations preserve exactly the
same class of formulas, which the theorem also characterizes
by deletion-monotonicity.

Note also that taking $\psi=\varphi$ in (8) of the first list gives $\models\varphi\to[\varphi]\varphi$.
Hence formulas equivalent to universal formulas are successful in PAL and satisfy the analogous success condition in BPAL (see Definition~\ref{def:four-notions}).
In BPAL, taking $\psi=\varphi$ in (8) of the second list and using announcement duality yields
$\models\neg\varphi\to[\uparrow\varphi]\neg\varphi$.
Thus formulas equivalent to existential formulas are impossible lies.
In PAL, the self-announcement instance $\models\langle!\varphi\rangle\varphi\to\varphi$ is valid for every formula, since the announcement is executable only where $\varphi$ is true.
\begin{corollary}[Propositional commutation]\label{cor:propositional-commutation}
  Let $\varphi$ be a formula of the relevant announcement language. Then the following are equivalent over K, both in PAL and BPAL:
  \begin{enumerate}
    \item $\varphi$ is equivalent to a propositional formula.
    \item $\varphi$ is equivalent to a universal formula and to an existential formula.
    \item $U_\varphi\circ U_\psi=U_{\varphi\land\psi}$ for all formulas $\psi$.
    \item $U_p\circ U_\varphi=U_\varphi\circ U_p$ for all proposition letters $p$.
  \end{enumerate}
\end{corollary}
\begin{proof}
Every propositional formula has a negation normal form belonging to both fragments, so (1) implies (2).
The equivalence $(2)\Longleftrightarrow(3)$ follows directly from the two operator comparisons in Theorem~\ref{thm:monotonicity}.

To prove $(2)\Rightarrow(1)$, assume (2).
By Theorem~\ref{thm:monotonicity}, $\varphi$ is both deletion-monotone and deletion-antimonotone.
Reduce $\varphi$ to an equivalent static formula $\theta\in\mathcal L$.
For any $M=(W,\{R_i\}_{i\in G},V)$, let $M^0=(W,\{\varnothing\}_{i\in G},V)$.
The two preservation properties give
\[
M,w\models\varphi
\quad\Longleftrightarrow\quad
M^0,w\models\theta.
\]
Let $\theta^0$ be obtained recursively from $\theta$ by keeping atoms and Boolean connectives and replacing every subformula of the form $\Box_i\chi$ by $\top$.
Induction on $\theta$ gives $M^0,w\models\theta$ iff $M,w\models\theta^0$.
Thus $\varphi$ is equivalent to the propositional formula $\theta^0$, proving (1).

$(1)\Rightarrow(4)$ is trivial. 
For $(4)\Rightarrow (2)$, assume (4), reduce $\varphi$ to an equivalent static formula, and choose a proposition letter $p$ not occurring in it; by (4), $U_p$ and $U_\varphi$ commute.
Given $N=M\upharpoonright S$ and $w\in S$, set $V(p)=S$.
In PAL, membership of $w$ in $U_p(U_\varphi(M))$ is equivalent to $M,w\models\varphi$, whereas membership in $U_\varphi(U_p(M))$ is equivalent to $N,w\models\varphi$.
Commutation with $U_p$ therefore equates these two truth values.
In BPAL, add a new state $z$ with $p$ false there and only the new arrow $zR_iw$, for some $i\in G$, obtaining $M^+$.
No old state can reach $z$, so truth of the static formula at old states is unchanged.
Survival of $(z,w)$ in $U_p(U_\varphi(M^+))$ is equivalent to $M,w\models\varphi$; survival in $U_\varphi(U_p(M^+))$ is equivalent to $N,w\models\varphi$, since $U_p(M^+),w$ and $N,w$ have the same generated submodel.
Again, commutation equates the two truth values.
Thus $\varphi$ is preserved under both submodels and model extensions, and Theorem~\ref{thm:monotonicity} yields (2).
\end{proof}
\subsection{Deflationarity and idempotence}\label{subsec:deflationarity-and-idempotence}
Next, we analyze deflationarity and idempotence. Recall that a formula $\varphi$ is successful iff $\models\varphi\to[\varphi]\varphi$ and self-refuting iff $\models\varphi\to[\varphi]\lnot\varphi$ (Definition~\ref{def:four-notions}). We will see that these notions are characterized using $U_\varphi$ as in the following proposition.
\begin{proposition}[Deflationarity and idempotence]\label{prop:deflationarity-and-idempotence}
  Let $\varphi$ be a formula of the relevant announcement language. Over K, both in PAL and BPAL:
  \begin{enumerate}
  \item $U_\varphi$ is deflationary: $U_\varphi(M)\preceq M$.
  \item $U_\varphi$ satisfies idempotence ($U_\varphi^2=U_\varphi$) iff $\varphi$ is successful.
  \item $U_\varphi^2=U_\bot$ iff $\varphi$ is self-refuting.
  \item If $\varphi$ is universal, $U_\varphi$ is idempotent.
  \end{enumerate}
\end{proposition}
\begin{proof}
\emph{(1).}
In PAL, $U_\varphi(M)$ is the induced submodel on $\llbracket\varphi\rrbracket_M$, so $U_\varphi(M)\preceq_P M$.
In BPAL, the domain and valuation are unchanged, and
\[
R_i^{U_\varphi(M)}
=R_i^M\cap\bigl(W\times\llbracket\varphi\rrbracket_M\bigr)
\subseteq R_i^M
\qquad(i\in G).
\]
Thus $U_\varphi(M)\preceq_B M$ in this case as well.

\emph{(2).}
Put $\theta=\varphi\land[\varphi]\varphi$.
By Lemma~\ref{lem:sequential-announcements}, $U_\varphi^2=U_\theta$.
Lemma~\ref{lem:announcement-operators-and-validity} therefore gives
$U_\varphi^2=U_\varphi$ iff $\models\theta\leftrightarrow\varphi$ in PAL,
and iff $\models\Box_i(\theta\leftrightarrow\varphi)$ for every $i\in G$ in BPAL.
Over K, $\models\Box_i\delta$ iff $\models\delta$:
if $\delta$ fails at $M,w$, add a fresh state $z$ with an $i$-arrow to $w$ and no arrows from old states to $z$.
Truth at old states is unchanged, also for announcement formulas by reduction to the static language, so $\Box_i\delta$ fails at $z$.
The other direction follows immediately from the semantics.
Since $G$ is nonempty, in either logic we obtain
\[
U_\varphi^2=U_\varphi
\quad\Longleftrightarrow\quad
\models(\varphi\land[\varphi]\varphi)\leftrightarrow\varphi
\quad\Longleftrightarrow\quad
\models\varphi\to[\varphi]\varphi.
\]
The last condition is precisely success.

\emph{(3).}
With $\theta$ as in (2), the same argument applied to $U_\bot$ yields
\[
U_\varphi^2=U_\bot
\quad\Longleftrightarrow\quad
\models\neg(\varphi\land[\varphi]\varphi)
\quad\Longleftrightarrow\quad
\models\varphi\to[\varphi]\neg\varphi.
\]
For the last equivalence, under the assumption $\varphi$, the announcement is executable and
$\neg[\varphi]\varphi$ is equivalent to $[\varphi]\neg\varphi$ in both logics.
Thus the last condition is exactly self-refutation.

\emph{(4).}
If $\varphi$ is universal, the universal part of Theorem~\ref{thm:monotonicity}, condition (8), with $\psi=\varphi$ gives
$\models\varphi\to[\varphi]\varphi$.
Hence $\varphi$ is successful, and (2) gives $U_\varphi^2=U_\varphi$.
\end{proof}

\section{Dynamic properties}\label{sec:dynamic-properties}
In this section, we investigate the dynamic properties of the public announcement operators listed in Table~\ref{tab:announcement-operator-properties}.
\begin{table}[htbp]
\centering
\caption{Basic properties of announcement operators.}
\label{tab:announcement-operator-properties}
\setlength{\tabcolsep}{4.5pt}
\renewcommand{\arraystretch}{1.15}
\begin{tabular}{p{0.20\textwidth} p{0.34\textwidth} p{0.38\textwidth}}
\hline
\textbf{Property} & \textbf{Definition} & \textbf{Interpretation} \\
\hline

\raggedright Fixed point
&
\(U_\varphi(M)=M\)
&
\raggedright The model $M$ is unchanged by the announcement of $\varphi$.
\tabularnewline[0.8em]

\raggedright Stabilization of $M$
&
\(\begin{gathered}[t]
U_\varphi^{\alpha+1}(M)=U_\varphi^{\alpha}(M)\\
\text{for some ordinal }\alpha
\end{gathered}\)
&
\raggedright The orbit of $M$ becomes constant after some ordinal stage.
\tabularnewline[0.8em]

\raggedright Stabilization rank
&
\(\begin{gathered}[t]
\rho_\varphi(M)=\min\{n<\omega:\\
U_\varphi^{n+1}(M)=U_\varphi^n(M)\},\\[0.4em]
\widehat\rho_\varphi(M)=\min\{\alpha:\\
U_\varphi^{\alpha+1}(M)=U_\varphi^\alpha(M)\}.
\end{gathered}\)
&
\raggedright The least finite or transfinite stage at which the orbit of $M$ under $U_\varphi$ reaches a fixed point.
\tabularnewline[0.4em]

\hline
\end{tabular}
\end{table}

\subsection{Fixed points}
Let $\mathcal{C}$ be the class of models, including the empty model in PAL.
For $F\colon \mathcal{C}\to\mathcal{C}$, let
\[
\operatorname{Fix}(F)=\{M\in\mathcal{C}\colon F(M)=M\}
\]
be the class of fixed points of $F$.

\begin{definition}[Stabilization rank]\label{def:stabilization-rank}
Let $F\colon\mathcal{C}\to\mathcal{C}$ be such that $F(N)\preceq N$ for every model $N$.
In BPAL, additionally require $W^{F(N)}=W^N$ and $V^{F(N)}=V^N$ for every $N$, so that $F$ can only delete accessibility arrows.
For an initial model $M=(W,\{R_i\}_{i\in G},V)$ and each ordinal $\alpha$, define $F^\alpha(M)$ by
\[
F^0(M)=M,\qquad F^{\alpha+1}(M)=F\bigl(F^\alpha(M)\bigr),
\]
and for a limit ordinal $\lambda$,
\[
F^\lambda(M)=
\begin{cases}
M\upharpoonright\displaystyle\bigcap_{\beta<\lambda}W^{F^\beta(M)},
&\text{in PAL},\\[1.2em]
\left(W,\left\{\displaystyle\bigcap_{\beta<\lambda}
R_i^{F^\beta(M)}\right\}_{i\in G},V\right),
&\text{in BPAL}.
\end{cases}
\]
Define the \emph{finite stabilization rank} $\rho_F(M)$ and the (\emph{transfinite}) \emph{stabilization rank} $\widehat\rho_F(M)$ of a model $M$ under the update operation $F$, respectively, by
\[
\begin{aligned}
\rho_F(M)&=\min\{n<\omega:F^{n+1}(M)=F^n(M)\},\\
\widehat\rho_F(M)&=\min\{\alpha:F^{\alpha+1}(M)=F^\alpha(M)\}.
\end{aligned}
\]
If no finite stage is stable, put $\rho_F(M)=\infty$.
The ordinal rank always exists, as explained below.
Finally, put
\[
F^\infty(M):=F^{\widehat\rho_F(M)}(M).
\]
We abbreviate $\rho_{U_\varphi}$ and $\widehat\rho_{U_\varphi}$ by
$\rho_\varphi$ and $\widehat\rho_\varphi$.
\end{definition}

Every strict successor step removes a state in PAL or a labelled arrow $(i,w,v)$ in BPAL, and no removed item can return.
There are only set-many original states or labelled arrows, so a strict decrease at every ordinal stage is impossible.
Hence $\widehat\rho_F(M)$ exists.
Once $F^{\alpha+1}(M)=F^\alpha(M)$, all subsequent iterates are equal, including at limit stages, by the defining intersections.
If $\rho_F(M)<\infty$, then $F^\infty(M)=F^{\rho_F(M)}(M)$.

\begin{theorem}[Fixed points]\label{thm:fixed-points}
  Let $\varphi$ and $\psi$ be formulas.
  \begin{enumerate} 
    \item In PAL:\[U_\varphi(M)=M\iff M\models\varphi\]
    and in BPAL:
    \[U_\varphi(M)=M\iff M\models\bigwedge_{i\in G}\Box_i\varphi\]
    \item 
    \[
\operatorname{Fix}(U_{\neg\varphi})
=
\{M\in\mathcal C:U_\varphi(M)=U_\bot(M)\}.
\]
\[
\operatorname{Fix}(U_\psi\circ U_\varphi)
=\operatorname{Fix}(U_\varphi\circ U_\psi)
=\operatorname{Fix}(U_{\varphi\land\psi})
=\operatorname{Fix}(U_\varphi)\cap
  \operatorname{Fix}(U_\psi)
\]
\[\operatorname{Fix}(U_\varphi)\cup\operatorname{Fix}(U_\psi)\subseteq \operatorname{Fix}(U_{\varphi\lor\psi})
\]
The last inclusion can be strict. 
\item If $\varphi$ is universal, we have \[
U_\varphi(M)=M,\quad N\preceq M
\quad\Longrightarrow\quad
U_\varphi(N)=N
\]
\[
U_\varphi^\infty(M)=U_\varphi(M)
\]
\[
U_\varphi\circ U_\psi\circ U_\varphi
=
U_\psi\circ U_\varphi.
\]
If $\psi$ is universal as well, then
\[
(U_\psi \circ U_\varphi)^\infty=U_\psi \circ U_\varphi\text{ and } (U_\varphi\circ  U_\psi)^\infty=U_\varphi\circ U_\psi
\]
but 
$(U_\psi\circ U_\varphi)^\infty=(U_\varphi\circ U_\psi)^\infty$ does not hold in general.
\item If $\varphi$ is existential, we have 
\[
U_\varphi(N)=N,\quad N\preceq M
\quad\Longrightarrow\quad
N\preceq U_\varphi(M)\preceq M.
\]
\[
U_\varphi^\infty(M)
=
\bigvee_{\substack{N\preceq M\\U_\varphi(N)=N}}N
\]
If $\psi$ is existential as well, then
\[
(U_\psi\circ U_\varphi)^\infty
=
(U_\varphi\circ U_\psi)^\infty
=
U_{\varphi\land\psi}^\infty.
\]
  \end{enumerate}
\end{theorem}
\begin{proof}
Every announcement operator is deflationary by Proposition~\ref{prop:deflationarity-and-idempotence}.

\emph{(1).}
In PAL, $U_\varphi(M)=M$ exactly when every state of $M$ satisfies $\varphi$.
In BPAL, $U_\varphi(M)=M$ exactly when the target of every accessibility arrow in $M$ satisfies $\varphi$.
These are the two stated global truth conditions, respectively.

\emph{(2).}
In PAL, $U_{\neg\varphi}(M)=M$ exactly when $\llbracket\varphi\rrbracket_M=\varnothing$, which is equivalent to $U_\varphi(M)=U_\bot(M)$.
In BPAL, $U_{\neg\varphi}(M)=M$ exactly when no accessibility arrow has a target satisfying $\varphi$.
This means that $U_\varphi$ deletes all arrows, again giving $U_\varphi(M)=U_\bot(M)$.

If $U_\psi(U_\varphi(M))=M$, then deflationarity gives
\[
M=U_\psi(U_\varphi(M))\preceq U_\varphi(M)\preceq M.
\]
Thus $U_\varphi(M)=M$ and then $U_\psi(M)=M$; the converse is immediate.
Interchanging $U_\varphi$ and $U_\psi$ proves the assertions about both compositions.
By (1), being fixed by $U_{\varphi\land\psi}$ is likewise equivalent to being fixed by both $U_\varphi$ and $U_\psi$.
Also by (1), every fixed point of either $U_\varphi$ or $U_\psi$ is fixed by $U_{\varphi\lor\psi}$.

For strictness, take $W=\{a,b\}$, $R_j=W\times W$ for every $j\in G$, and $V(p)=\{a\}$.
With $\varphi=p$ and $\psi=\neg p$, both $U_p$ and $U_{\neg p}$ change this model in either logic, whereas $U_{p\lor\neg p}(M)=M$.
Hence the inclusion can be strict.

\emph{(3).}
Suppose that $\varphi$ is universal.
By (1) and Theorem~\ref{thm:monotonicity}, $U_\varphi(M)=M$ and $N\preceq M$ imply $U_\varphi(N)=N$.
For BPAL, apply preservation under induced submodels and arrow deletion to the universal formula $\bigwedge_{i\in G}\Box_i\varphi$.
By Proposition~\ref{prop:deflationarity-and-idempotence}, $U_\varphi^2=U_\varphi$, so $U_\varphi^\infty=U_\varphi$.
Since $U_\varphi(M)$ is fixed by $U_\varphi$ and $U_\psi(U_\varphi(M))\preceq U_\varphi(M)$, the first assertion gives $U_\varphi(U_\psi(U_\varphi(M)))=U_\psi(U_\varphi(M))$.
This proves the absorption identity $U_\varphi\circ U_\psi\circ U_\varphi=U_\psi\circ U_\varphi$ for arbitrary $\psi$.
If $\psi$ is universal as well, then $U_\psi^2=U_\psi$, and consequently
\[
(U_\psi\circ U_\varphi)^2
=U_\psi\circ(U_\varphi\circ U_\psi\circ U_\varphi)
=U_\psi^2\circ U_\varphi
=U_\psi\circ U_\varphi.
\]
Interchanging $U_\varphi$ and $U_\psi$ proves the other stabilization identity. $(U_\psi\circ U_\varphi)^\infty=(U_\varphi\circ U_\psi)^\infty$ does not hold in general as we have already seen for the universal formulas $\varphi=p$ and $\psi=\Box_i p$ at the beginning of Section~\ref{sec:algebraic-properties}. 

\emph{(4).}
Suppose that $\varphi$ is existential.
Then $U_\varphi$ is monotone by Theorem~\ref{thm:monotonicity}.
If $U_\varphi(N)=N$ and $N\preceq M$, monotonicity and deflationarity give
\[
N=U_\varphi(N)\preceq U_\varphi(M)\preceq M.
\]
More generally, let $T$ be any monotone model operator satisfying the assumptions of Definition~\ref{def:stabilization-rank}.
For every fixed point $N\preceq M$ of $T$, transfinite induction gives
\[
N\preceq T^\alpha(M)\qquad\text{for every ordinal }\alpha.
\]
The successor step uses $N=T(N)\preceq T(T^\alpha(M))$; at a limit stage, the inclusion persists under the defining intersection.
Since $T^\infty(M)$ is itself a fixed point below $M$, it is therefore the greatest such fixed point.
Applying this observation to $T=U_\varphi$ yields the stated equality for $U_\varphi^\infty(M)$.

Finally, if $\psi$ is existential as well, the operators $U_\psi\circ U_\varphi$, $U_\varphi\circ U_\psi$, and $U_{\varphi\land\psi}$ are all monotone and deflationary, and preserve the domain and valuation in BPAL.
By (2), they have the same fixed points.
The preceding argument identifies each stabilized value at $M$ with the greatest common fixed point below $M$, proving the final equalities.
\end{proof}
\subsection{Stabilization}\label{subsec:stabilization}Next, we examine stabilization.
We write $U_\varphi^P$ and $U_\varphi^B$ for the PAL and BPAL operators, respectively. Also, we say that a model $M$ is \emph{image-finite} if each state has finitely many $i$-successors for every $i\in G$.
\begin{theorem}[Stabilization rank]\label{thm:stabilization-rank}
Let $\varphi\in\mathcal{L}$.
  \begin{enumerate}
    \item For all models $M$, \[\widehat{\rho}_\varphi^{B}(M)
\leq
\widehat\rho_\varphi^{P}(M)
\leq
\widehat\rho_\varphi^{B}(M)+1.\] If each state has an incoming arrow (e.g., in S5),
\[\widehat\rho_\varphi^{P}(M)=\widehat\rho_\varphi^{B}(M).\]
\item If $M$ is finite, $\widehat{\rho}_\varphi(M)\leq |W|$.
\item If $M$ is image-finite, $\widehat{\rho}_\varphi(M)\leq \omega$. 
\item 
There exists a formula $\varphi\in\mathcal L$ such that, for each ordinal $\alpha>0$, there is a model $M_\alpha$ satisfying \[\widehat{\rho}_\varphi^P(M_\alpha)=\widehat{\rho}_\varphi^B(M_\alpha)=\alpha.\]
If $|G|\geq 2$, the models $M_\alpha$ can all be chosen to be S5 models.\footnote{The assumption $|G|\geq2$ is essential here.
If $G=\{i\}$ and $\varphi$ contains $k$ proposition letters, every S5 model $M$ satisfies $\widehat\rho_\varphi^P(M)=\widehat\rho_\varphi^B(M)\leq2^k$.
Within each initial equivalence class, states agreeing on these letters agree on $\varphi$ at every PAL stage; hence each strict change in that class removes at least one of its at most $2^k$ valuation types.
Once a class stops changing, it remains fixed.
Thus all classes stabilize within $2^k$ steps, and (1) gives the BPAL bound.}
\end{enumerate}
\end{theorem}
\begin{proof}
\emph{(1).}
Fix $M=(W,\{R_i\}_{i\in G},V)$, and put
\[
P_\alpha=(U_\varphi^P)^\alpha(M),\qquad
B_\alpha=(U_\varphi^B)^\alpha(M),\qquad
S_\alpha=W^{P_\alpha}.
\]
We first prove, by transfinite induction, the orbit correspondence
\begin{equation}\label{eq:pal-bpal-orbit-correspondence}
P_\alpha=M\upharpoonright S_\alpha,\qquad
B_\alpha=\bigl(W,\{R_i\cap(W\times S_\alpha)\}_{i\in G},V\bigr).
\end{equation}
The first equality follows directly from the definition of PAL iteration, and both equalities hold at stage $0$.
Suppose that the correspondence holds at $\alpha$.
At each $w\in S_\alpha$, the generated submodels of $P_\alpha$ and $B_\alpha$ coincide: all successors in either model remain in $S_\alpha$, with the same accessibility relations and valuations there.
Thus these models agree at $w$ on every formula in $\mathcal L$, and in particular
\begin{equation}\label{eq:pal-survivor-recurrence}
S_{\alpha+1}=S_\alpha\cap\llbracket\varphi\rrbracket_{B_\alpha}.
\end{equation}
The $i$-relation of $B_{\alpha+1}$ is consequently
\[
\bigl(R_i\cap(W\times S_\alpha)\bigr)
\cap\bigl(W\times\llbracket\varphi\rrbracket_{B_\alpha}\bigr)
=R_i\cap(W\times S_{\alpha+1}),
\]
proving the successor step.
At a nonzero limit $\lambda$, we have $S_\lambda=\bigcap_{\beta<\lambda}S_\beta$, and
\[
\bigcap_{\beta<\lambda}\bigl(R_i\cap(W\times S_\beta)\bigr)
=R_i\cap(W\times S_\lambda),
\]
which proves the limit step.

If $P_{\alpha+1}=P_\alpha$, then $S_{\alpha+1}=S_\alpha$, so the correspondence gives $B_{\alpha+1}=B_\alpha$.
Hence $\widehat\rho_\varphi^B(M)\leq\widehat\rho_\varphi^P(M)$.
Conversely, let $\beta=\widehat\rho_\varphi^B(M)$.
Since $B_{\beta+1}=B_\beta$, equation~\eqref{eq:pal-survivor-recurrence} gives
\[
S_{\beta+2}
=S_{\beta+1}\cap\llbracket\varphi\rrbracket_{B_{\beta+1}}
=S_{\beta+1}\cap\llbracket\varphi\rrbracket_{B_\beta}
=S_{\beta+1}.
\]
Thus $\widehat\rho_\varphi^P(M)\leq\beta+1$.
If every state of the initial model has an incoming arrow, then $B_{\beta+1}=B_\beta$ already forces $S_{\beta+1}=S_\beta$.
Indeed, a state $v\in S_\beta\setminus S_{\beta+1}$ would have an original incoming arrow $wR_iv$, which belongs to $B_\beta$ but not to $B_{\beta+1}$ by~\eqref{eq:pal-bpal-orbit-correspondence}.
This proves equality of the ranks under the stated assumption.

\emph{(2).}
In PAL, each strict update removes at least one state, and a removed state never returns.
There can therefore be at most $|W|$ strict updates when $W$ is finite.
The BPAL bound follows from (1).

\emph{(3).}
Suppose that $M$ is image-finite.
Let $d$ be the modal depth of $\varphi$, and retain the notation from (1).
For $w\in S_\omega$, let $D_d(w)$ be the set of states reachable from $w$ in $M$ by a path of length at most $d$, using any of the relations $R_i$.
Since $G$ is finite and $M$ is image-finite, $D_d(w)$ is finite.
The decreasing sequence $S_n\cap D_d(w)$ therefore eventually equals $S_\omega\cap D_d(w)$: choose $n_w<\omega$ after every state of $D_d(w)\setminus S_\omega$ has been removed.
For $n\geq n_w$, the induced models $P_n$ and $P_\omega$ consequently have the same surviving states, relations, and valuations within this depth-$d$ neighborhood of $w$.
Induction on formulas of modal depth at most $d$ gives
\[
P_n,w\models\varphi\quad\Longleftrightarrow\quad P_\omega,w\models\varphi.
\]
But $w\in S_{n+1}$ for every $n<\omega$, so $P_n,w\models\varphi$ for every $n$.
It follows that every $w\in S_\omega$ satisfies $\varphi$ in $P_\omega$.
Thus $P_{\omega+1}=P_\omega$, also when $S_\omega$ is empty, and the BPAL bound again follows from (1).

\emph{(4).}
First suppose that $G=\{i\}$, and put $\varphi:=p\land\Diamond_i(p\land\Box_i\neg p)$.
For an ordinal $\alpha>0$, let $W_\alpha=\alpha\cup\{z\}$, where $z\notin\alpha$, set $V(p)=\alpha$, and give all other proposition letters the empty valuation.
Among ordinal states, put $\gamma R_i\eta$ iff $\eta<\gamma$; also put $zR_iw$ for every $w\in W_\alpha$, including $w=z$.
There are no other arrows.
Let $M_\alpha$ be the resulting model and $S_\beta$ the domain after $\beta$ PAL announcements of $\varphi$.
Initially, $z$ falsifies $p$ and is therefore removed at the first step.
For each $\beta<\alpha$, if the surviving ordinal states are precisely those at least $\beta$, then $\beta$ is the only one satisfying $p\land\Box_i\neg p$, so $\varphi$ holds exactly at the ordinals greater than $\beta$.
Together with intersection at limit stages, this proves by transfinite induction that $S_\beta=\{\gamma<\alpha:\gamma\geq\beta\}$ for $0<\beta\leq\alpha$.
Every $\beta<\alpha$ is removed at stage $\beta+1$, and $S_\alpha=\varnothing$, so $\widehat\rho_\varphi^P(M_\alpha)=\alpha$.
Every initial state has an incoming arrow from $z$, hence (1) gives $\widehat\rho_\varphi^B(M_\alpha)=\alpha$ as well.

Now suppose that $|G|\geq2$, and fix distinct agents $a,b\in G$ and a proposition letter $p$.
For any formula $\chi$, abbreviate 
\[L\chi:=(p\land\Diamond_a(\neg p\land\chi))\lor(\neg p\land\Diamond_b(p\land\chi)),
\]
 and put $\delta:=L\top$ and $\varphi:=L\neg\delta$.
Now fix an ordinal $\alpha>0$, and let $W_\alpha$ be the set of nonempty finite strictly decreasing sequences of ordinals below $\alpha$.
For $s\in W_\alpha$, write $\ell(s)$ for its length, $\operatorname{last}(s)$ for its last entry, and 
\[C(s):=\{s^\frown(\gamma):\gamma<\operatorname{last}(s)\}
\] 
for its set of immediate children.
Set $V(p)=\{s\in W_\alpha:\ell(s)\text{ is odd}\}$, and give all other proposition letters the empty valuation.
Define $R_a$ by the equivalence classes $\{s\}\cup C(s)$ for all odd-length $s$.
Define $R_b$ by the equivalence classes $\{s\}\cup C(s)$ for all even-length $s$, together with singleton classes for all length-one sequences.
These are partitions: each non-root sequence has a unique immediate predecessor of the opposite length parity.
For any other agent, use the identity relation.
The resulting model $M_\alpha$ is therefore S5.

For any $S\subseteq W_\alpha$ and $s\in S$, the modality selected by the truth value of $p$ at $s$ sees precisely its own class $\{s\}\cup C(s)$, restricted to $S$.
The required opposite truth value of $p$ excludes $s$ and holds at every child.
Thus $L\chi$ holds at $s$ exactly when some $t\in C(s)\cap S$ satisfies $\chi$ in $M_\alpha\upharpoonright S$.
In particular, $M_\alpha\upharpoonright S,s\models\delta$ iff $C(s)\cap S\ne\varnothing$.

Let $S_\beta$ denote the domain after $\beta$ PAL announcements of $\varphi$ in $M_\alpha$.
We prove by transfinite induction that $S_\beta=\{s\in W_\alpha:\operatorname{last}(s)\geq\beta\}$ for every $\beta\leq\alpha$.
The base case is immediate.
For the successor step, assume $\beta<\alpha$ and take $s\in S_\beta$.
If $\operatorname{last}(s)=\beta$, no child of $s$ remains in $S_\beta$, so $\delta$ and hence $\varphi$ are false at $s$.
If $\operatorname{last}(s)>\beta$, the child $t=s^\frown(\beta)$ belongs to $S_\beta$, but none of its children do.
Thus $\neg\delta$ holds at $t$, and $\varphi=L\neg\delta$ holds at $s$.
Consequently, exactly the states with last entry greater than $\beta$ survive the next announcement, proving the successor step.
At a limit stage, the claim follows by intersection.
For every $\beta<\alpha$, the length-one sequence $(\beta)$ is removed at stage $\beta+1$, whereas $S_\alpha=\varnothing$.
Hence $\widehat\rho_\varphi^P(M_\alpha)=\alpha$, and reflexivity together with (1) yields $\widehat\rho_\varphi^B(M_\alpha)=\alpha$.
\end{proof}
\begin{figure}[htbp]
    \centering
    \resizebox{0.98\textwidth}{!}{%
    \begin{tikzpicture}[
        x=1cm, y=1cm,
        >=Stealth,
        line cap=round,
        line join=round,
        every node/.style={font=\small},
        world/.style={
            circle, fill=black, inner sep=0pt,
            minimum size=4pt
        },
        iedge/.style={->, black, thick},
        announcement/.pic={
            \node[font=\Large, inner sep=0pt] {$\Longrightarrow$};
        },
        state/.style={
            circle, draw, fill=white,
            minimum size=8mm, inner sep=1pt,
            font=\small
        },
        aedge/.style={<->, blue, thick},
        bedge/.style={<->, red, thick},
        asib/.style={blue, thick},
        bsib/.style={red, thick},
        omission/.style={fill=white, inner sep=1.5pt}
    ]

    % (a) Single-agent construction and its PAL iterates.
    \node[anchor=west] at (-5.65,1.4)
        {(a) Single agent (PAL)};

    % Initial ordinal states.
    \node[world, label=above:$0$, label=below:$\neg\varphi$]
        (s0) at (-5.2,0) {};
    \node[world, label=above:$1$, label=below:$\varphi$]
        (s1) at (-4.25,0) {};
    \node[world, label=above:$2$, label=below:$\varphi$]
        (s2) at (-3.3,0) {};
    \node at (-2.4,0) {$\cdots$};
    \node at (-1.4,0) {$<\alpha$};

    \draw[iedge] (s1) -- (s0);
    \draw[iedge] (s2) -- (s1);
    \draw[iedge] (-2.7,0) -- (s2);

    % The auxiliary state z sees every initial state, including itself.
    \node[world, label=above right:$z$, label=right:$\neg p$]
        (z) at (-2.2,-1.3) {};
    \draw[iedge] (z) -- (s0);
    \draw[iedge] (z) -- (s1);
    \draw[iedge] (z) -- (s2);
    \draw[iedge] (z) edge[loop below, min distance=8mm] (z);
    \node at (-2.2,-2.2) {$\neg\varphi$};

    \node at (-3.7,-2.65) {$M_\alpha$};

    % Repeated announcements.
    \pic at (-0.35,0) {announcement};
    \node at (-0.35,0.48) {$\varphi$};
    \node at (0.475,0) {$\cdots$};
    \pic at (1.3,0) {announcement};
    \node at (1.3,0.48) {$\varphi$};

    % At a positive stage beta, z and all ordinals below beta are gone.
    \node[world, label=above:$\beta$, label=below:$\neg\varphi$]
        (t0) at (2.7,0) {};
    \node[world, label=above:$\beta+1$, label=below:$\varphi$]
        (t1) at (3.9,0) {};
    \node[world, label=above:$\beta+2$, label=below:$\varphi$]
        (t2) at (5.1,0) {};
    \node at (6.1,0) {$\cdots$};
    \node at (7.1,0) {$<\alpha$};

    \draw[iedge] (t1) -- (t0);
    \draw[iedge] (t2) -- (t1);
    \draw[iedge] (5.8,0) -- (t2);

    % Every displayed ordinal state satisfies p.
    \foreach \x in {-5.2,-4.25,-3.3,2.7,3.9,5.1} {
        \node at (\x,0.75) {$p$};
    }

    \node at (4.9,-2.65)
        {$\bigl(U_\varphi^P\bigr)^\beta(M_\alpha)
          \qquad(0<\beta<\alpha)$};

    % (b) S5 construction.
    \begin{scope}[yshift=-4.2cm]

        \node[anchor=west] at (-5.65,0.9)
            {(b) At least two agents (S5)};

        % Length-one sequences: distinct connected components.
        \node[state] (n0) at (-5.2,0)
            {$\langle 0\rangle$};
        \node[state] (n1) at (-3.7,0)
            {$\langle 1\rangle$};
        \node[state] (n2) at (-1.4,0)
            {$\langle 2\rangle$};
        \node at (0.55,0) {$\cdots$};
        \node[state] (ng) at (3.2,0)
            {$\langle\gamma\rangle$};
        \node at (6.7,0) {$\cdots$};

        \node at (3.2,0.9)
            {$0<\eta<\beta<\gamma<\alpha$};

        % Length-two sequences.
        \node[state] (n10) at (-3.7,-2.1)
            {$\langle 1,0\rangle$};
        \node[state] (n20) at (-2.25,-2.1)
            {$\langle 2,0\rangle$};
        \node[state] (n21) at (-0.55,-2.1)
            {$\langle 2,1\rangle$};
        \node[state] (ng0) at (1.55,-2.1)
            {$\langle\gamma,0\rangle$};
        \node[state] (ngb) at (4.75,-2.1)
            {$\langle\gamma,\beta\rangle$};

        % Length-three sequences.
        \node[state] (n210) at (-0.55,-4.2)
            {$\langle 2,1,0\rangle$};
        \node[state] (ngb0) at (3.25,-4.2)
            {$\langle\gamma,\beta,0\rangle$};
        \node[state] (ngbe) at (5.85,-4.2)
            {$\langle\gamma,\beta,\eta\rangle$};

        % Length-four sequences.
        \node[state] (ngbe0) at (5.85,-6.3)
            {$\langle\gamma,\beta,\eta,0\rangle$};

        % R_a: blocks whose parent has odd length.
        \draw[aedge] (n1) -- (n10);
        \draw[aedge] (n2) -- (n20);
        \draw[aedge] (n2) -- (n21);
        \draw[aedge] (ng) -- (ng0);
        \draw[aedge] (ng) -- (ngb);
        \draw[aedge] (ngbe) -- (ngbe0);

        % R_b: blocks whose parent has even length.
        \draw[bedge] (n21) -- (n210);
        \draw[bedge] (ngb) -- (ngb0);
        \draw[bedge] (ngb) -- (ngbe);

        % Mutual accessibility among siblings in each block.
        \draw[asib] (n20) -- (n21);
        \draw[asib] (ng0) -- (ngb);
        \draw[asib] (ngb) -- (6.25,-2.1);
        \draw[bsib] (ngb0) -- (ngbe);
        \draw[bsib] (ngbe) -- (6.8,-4.2);
        \draw[asib] (ngbe0) -- (6.8,-6.3);

        % Omitted ordinal-indexed siblings and components.
        \node[omission] at (3.15,-2.1) {$\cdots$};
        \node[omission] at (6.7,-2.1) {$\cdots$};
        \node[omission] at (4.55,-4.2) {$\cdots$};
        \node[omission] at (7.05,-4.2) {$\cdots$};
        \node[omission] at (7.05,-6.3) {$\cdots$};

        % Valuation of p, determined by sequence-length parity.
        \node[right] at (7.85,0) {$p$};
        \node[right] at (7.85,-2.1) {$\neg p$};
        \node[right] at (7.85,-4.2) {$p$};
        \node[right] at (7.85,-6.3) {$\neg p$};

    \end{scope}
    \end{tikzpicture}%
    }
    \caption{Top: the single-agent model and its PAL iterate at stage
    $0<\beta<\alpha$. Black arrows denote $R_i$; some accessibility
    arrows are omitted.
    Bottom: the S5 model, with one component for each root
    $\langle\gamma\rangle$ ($\gamma<\alpha$).
    Blue and red bidirectional arrows denote $R_a$ and $R_b$;
    horizontal colored segments indicate mutual accessibility among siblings.
    Reflexive and further transitive arrows in the bottom panel are omitted.
    Only indices below $\alpha$ and valid decreasing sequences are included.}
    \label{fig:ordinal-rank-peeling-model}
\end{figure}
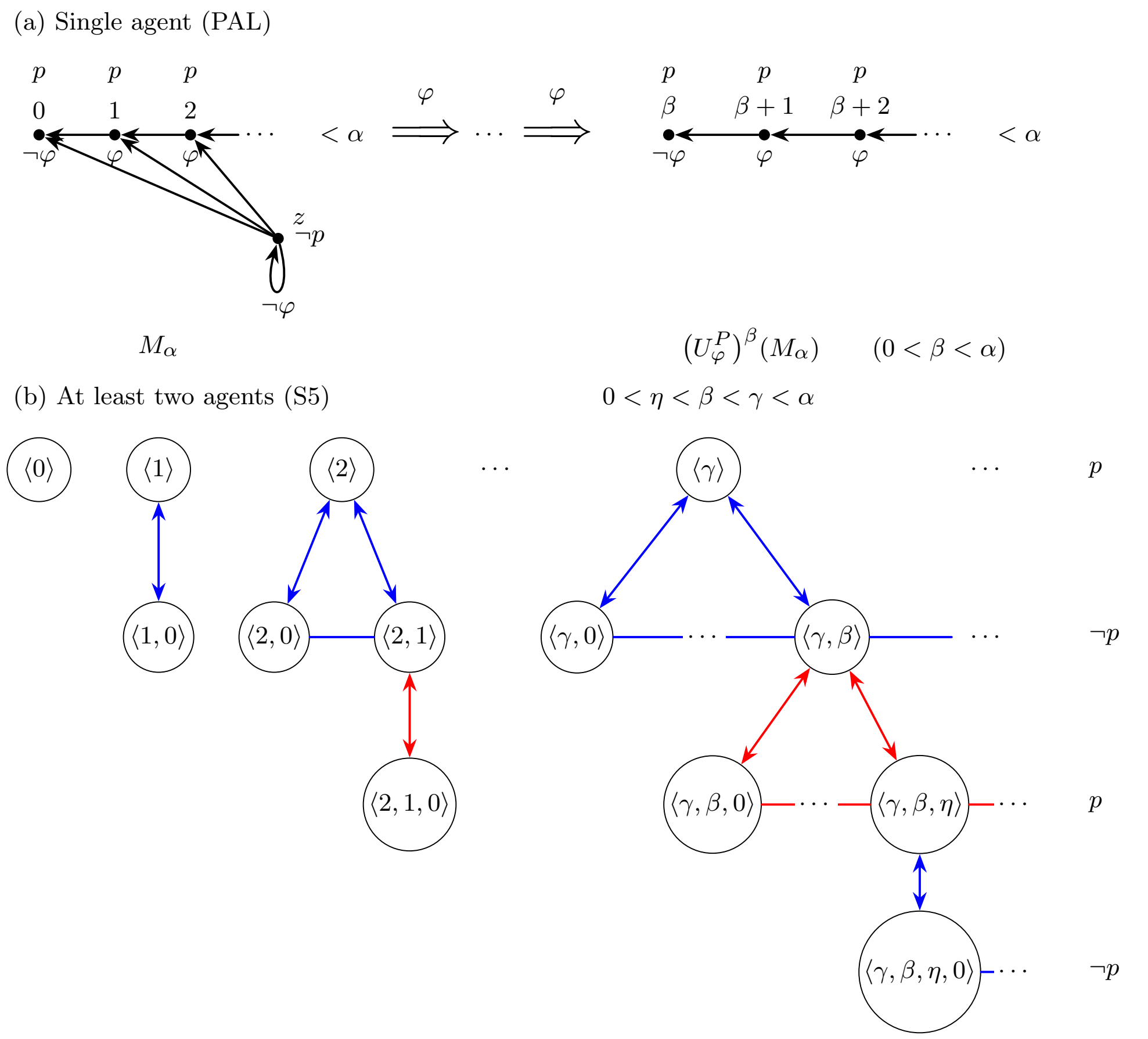

Theorem~\ref{thm:stabilization-rank}(1) says that, for $\varphi\in\mathcal L$, stabilization in BPAL is faster than or as fast as stabilization in PAL but the difference in stabilization rank is at most one. Part (4) says that, for some formula $\varphi$, every nonzero ordinal occurs as a stabilization rank under iterated announcements of $\varphi$; when $|G|\geq 2$, the witnessing models can be chosen to be S5. The constructions are illustrated in Figure~\ref{fig:ordinal-rank-peeling-model}. For condition (4), see also the remark on always informativeness right after Theorem~\ref{thm:extinction-under-iterated-announcements}. 

For formulas $\psi,\alpha\in\mathcal{L}$, define the relativization $\psi^\alpha$ of $\psi$ by $\alpha$ as
\begin{gather*}
  p^\alpha=p,\qquad
  (\lnot\psi)^\alpha=\lnot\psi^\alpha,\qquad
  (\psi\land\chi)^\alpha=\psi^\alpha\land\chi^\alpha,\\
  (\Box_i\psi)^\alpha=\Box_i(\alpha\to\psi^\alpha),\qquad
  (\Diamond_i\psi)^\alpha=\Diamond_i(\alpha\land\psi^\alpha).
\end{gather*}
We then have
\[
M\upharpoonright\llbracket\alpha\rrbracket_M,w\models\psi
\iff
M,w\models\psi^\alpha
\]
whenever $M,w\models\alpha$.

In the following theorem, we ask exactly when every model stabilizes after finitely many announcements since only finitely many announcements seem to be physically feasible. However, transfinite iteration is not only a natural extension of finite iteration, but also may admit meaningful epistemic or communicative interpretations, a possibility we leave for future investigation. See also \citet[Sections 1.2, 2.4, 5.1–5.4, and 6.2]{vanBenthem2007}, where higher-order reasoning about players’ rationality motivates iterative strategy elimination, and transfinite approximation connects the resulting epistemic processes to fixed points characterizing game-theoretic solution concepts such as rationalizability.
\begin{theorem}[Finite stabilization]\label{thm:finite-stabilization}
For a formula $\varphi\in\mathcal{L}$, let
$\theta_0:=\top$ and $\theta_{n+1}:=\theta_n\land\varphi^{\theta_n}$. Then, in either PAL or BPAL, the following are equivalent:
\begin{enumerate}
  \item For all models $M$, $\widehat\rho_\varphi(M)<\omega$.
  \item There exists an $N\geq 1$ such that for all models $M$, $\widehat\rho_\varphi(M)\leq N$.
  \item There exists an $N\geq 1$ such that $\models\theta_N\leftrightarrow\theta_{N+1}$.
  \item There exists an $N\geq 1$ such that $U_\varphi^N$ is idempotent.
\end{enumerate}
\end{theorem}
\begin{proof}
Write $F=U_\varphi$, initially in either semantics.
We first prove the equivalence of (1), (2), and (4) directly for both logics.
Clearly, (2) implies (1).
For the converse, suppose that (1) holds but (2) fails.
For every integer $n\geq1$, choose a model $M_n$ with $\widehat\rho_\varphi(M_n)>n$, and replace the models by disjoint copies if necessary.
Let $M=\bigsqcup_{n\geq1}M_n$ be their disjoint union, with relations and valuations defined componentwise.
Truth of static modal formulas is evaluated within each component, so announcement updates act independently on the components.
Induction on $k<\omega$ therefore gives, in either semantics,
\[
F^k(M)=\bigsqcup_{n\geq1}F^k(M_n).
\]
For every $k<\omega$, choose $n\geq\max\{1,k\}$.
Since $\widehat\rho_\varphi(M_n)>n$, we have $F^{k+1}(M_n)\ne F^k(M_n)$; otherwise all later iterates of that component would already be equal.
Thus $F^{k+1}(M)\ne F^k(M)$ for every $k$, contradicting (1).
This proves (1)$\Rightarrow$(2).

For each $N\geq1$, condition (2) with bound $N$ says exactly that $F^{N+1}=F^N$.
It implies $F^{2N}=F^N$, which is idempotence of $F^N$.
Conversely, if $F^{2N}=F^N$, deflationarity gives, for every model $M$,
\[
F^{2N}(M)\preceq F^{N+1}(M)\preceq F^N(M).
\]
The outer terms are equal, so $F^{N+1}(M)=F^N(M)$.
Hence (2) and (4) are equivalent in either logic.

By Theorem~\ref{thm:stabilization-rank}(1), the existence of a uniform finite bound in (2) is the same for PAL and BPAL.
A PAL bound $N$ is also a BPAL bound, and a BPAL bound $N$ gives a PAL bound $N+1$.
It therefore remains to connect (3) to these conditions in PAL.

The relativization equivalence preceding the theorem yields, by induction on $n$,
\begin{equation}\label{eq:finite-iteration-relativization}
(U_\varphi^P)^n(M)
=M\upharpoonright\llbracket\theta_n\rrbracket_M
=U_{\theta_n}^P(M).
\end{equation}
The base case follows from $\theta_0=\top$.
For the induction step, a state survives the next update precisely when it satisfies $\theta_n$ in $M$ and $\varphi$ in the induced submodel on $\llbracket\theta_n\rrbracket_M$.
By relativization, this is equivalent to satisfying $\theta_n\land\varphi^{\theta_n}=\theta_{n+1}$ in $M$.
Relations and valuations are the corresponding restrictions of those of $M$, proving~\eqref{eq:finite-iteration-relativization}.
Consequently, for a given $N\geq1$, all PAL iterates stabilize by $N$ exactly when $\theta_N$ and $\theta_{N+1}$ have the same truth set in every model.
This proves the equivalence of PAL conditions (2) and (3), completing the proof.
\end{proof}
We next ask exactly when every model stabilizes by stage $\omega$ since $\omega$-iterated announcements are understood as the limit behavior of finitely iterated announcements. Surprisingly, $\omega$-stabilization can be characterized by idempotence and commutativity:
\begin{theorem}[$\omega$-stabilization]\label{thm:omega-stabilization}
Let $\varphi\in\mathcal L$.
The following conditions are equivalent in either PAL or BPAL:
\begin{enumerate}
  \item For every model $M$, $\widehat\rho_\varphi(M)\leq\omega$.
  \item $U_\varphi^\omega$ is idempotent:
  \[
  U_\varphi^\omega\circ U_\varphi^\omega=U_\varphi^\omega.
  \]
  \item $U_\varphi$ and $U_\varphi^\omega$ commute:
  \[
  U_\varphi\circ U_\varphi^\omega=U_\varphi^\omega\circ U_\varphi.
  \]
\end{enumerate}
Moreover, these conditions hold in PAL if and only if they hold in BPAL.\footnote{Allowing countable conjunctions, put $\Theta_\omega=\bigwedge_{n<\omega}\theta_n$, with $\theta_n$ from Theorem~\ref{thm:finite-stabilization}. Then (1) is equivalent to $\models\Theta_\omega\to\varphi^{\Theta_\omega}$; relativization is extended to this infinitary language. We give no proof here.}
\end{theorem}
\begin{proof}
In either semantics, deflationarity gives, for every model $M$,
\[
U_\varphi^\omega(U_\varphi(M))
=\bigcap_{n<\omega}U_\varphi^{n+1}(M)
=\bigcap_{n<\omega}U_\varphi^n(M)
=U_\varphi^\omega(M),
\]
where the intersections are taken as in Definition~\ref{def:stabilization-rank}.
Thus $U_\varphi^\omega\circ U_\varphi=U_\varphi^\omega$ always holds.
Since (1) is equivalent to $U_\varphi\circ U_\varphi^\omega=U_\varphi^\omega$, this proves (1)$\Leftrightarrow$(3).

If (1) holds, then $U_\varphi^\omega(M)$ is fixed by $U_\varphi$ for every $M$, so further iteration leaves it unchanged and (2) follows.
Conversely, if (2) holds, deflationarity yields
\[
U_\varphi^\omega(U_\varphi^\omega(M))
\preceq U_\varphi(U_\varphi^\omega(M))
\preceq U_\varphi^\omega(M).
\]
The outer terms are equal by (2), so $U_\varphi(U_\varphi^\omega(M))=U_\varphi^\omega(M)$, proving (1).

It remains to compare the two semantics.
A PAL bound $\omega$ is also a BPAL bound by Theorem~\ref{thm:stabilization-rank}(1).
For the converse, suppose that (1) fails in PAL.
Choose a model $M$ and a state $v\in S_\omega\setminus S_{\omega+1}$, where
$S_\alpha=W^{(U_\varphi^P)^\alpha(M)}$.
Fix $i\in G$, and form $M^+$ by adding a new state $z$, with arbitrary atomic valuation, and only the new arrow $zR_iv$.
No original state can reach $z$.
Locality of static modal formulas and transfinite induction therefore show that, at every stage, the PAL iterate of $M^+$ restricted to the original domain coincides with the corresponding PAL iterate of $M$.
By the orbit correspondence~\eqref{eq:pal-bpal-orbit-correspondence} applied to $M^+$, the arrow $zR_iv$ belongs to its BPAL iterate at stage $\omega$ but not at stage $\omega+1$.
Hence (1) also fails in BPAL, proving the claimed equivalence between the semantics.
\end{proof}

For the next theorem, let $C_G\psi$ denote common belief, interpreted by
\[
M,w\models C_G\psi
\quad\Longleftrightarrow\quad
M,v\models\psi\text{ for every }v
\text{ with }w\Bigl(\bigcup_{i\in G}R_i\Bigr)^+v,
\]
where $+$ denotes transitive closure along paths of positive length.
A model is \emph{edgeless} if all its accessibility relations are empty. 

The following notions are common knowledge/belief versions of unbelievability and unknowability of a formula $\varphi$ for agent $i$, which are defined as unsatisfiability of $\Box_i\varphi$ and $\varphi\land\Box_i\varphi$, respectively~\citep{yamada2026Sources}. For different definitions of unknowability, see \citet{Balbiani2008,edgington1985,vanBenthem2004,vanDitmarschHoekIliev2012,liuEtAl2022,heylenMoralesCarbonell2023}.
\begin{definition}\label{def:common-believability-knowability-and-informativeness}
Let $\varphi$ be a formula.
\begin{itemize}
\item $\varphi$ is \emph{commonly believable} iff $C_G\varphi$ is satisfiable.
\item $\varphi$ is \emph{commonly knowable} iff $\varphi\land C_G\varphi$ is satisfiable.
\end{itemize}
``\emph{Commonly unbelievable}'' and ``\emph{commonly unknowable}'' refer to their negations.
\begin{itemize}
  \item $\varphi$ is \emph{always informative when true} (\emph{when false}) iff for all pointed models $M,w$, $M,w\models\varphi$ (resp., $M,w\models\lnot\varphi$)
implies $(M|\varphi)_w\ne M_w$.
\item $\varphi$ is \emph{always informative} iff $\varphi$ is always informative both when true and false.
\end{itemize}
\end{definition}
Note that in any class of frames, $\varphi$ is always informative iff $\varphi$ is commonly unbelievable. On any reflexive frame, any formula is always informative when false.

As the last theorem of the paper, we ask exactly when iterated announcements of $\varphi$ make a model empty or edgeless. Unexpectedly, this is equivalent to common unknowability, always informativeness when true, and a fixed point view of the generalized Moore sentence $\varphi\land\lnot C_G\varphi$.
\begin{theorem}[Extinction under iterated announcements]\label{thm:extinction-under-iterated-announcements}
Let $\varphi\in\mathcal L$. The following conditions are equivalent over K in either PAL or BPAL:
\begin{enumerate}
  \item For every finite model $M$, there is an $n<\omega$ such that
  \[
  U_\varphi^n(M)=U_\bot(M).
  \]
  \item For every model $M$, there is an ordinal $\alpha$ such that
  \[
  U_\varphi^\alpha(M)=U_\bot(M).
  \]
  \item Every fixed point $N$ of $U_\varphi$ satisfies $N=U_\bot(N)$.
  \item $\varphi$ is commonly unknowable.
  \item $\varphi$ is always informative when true.
  \item $\models C_G\varphi\to\neg\varphi$.
  \item $\models\varphi\leftrightarrow(\varphi\land\neg C_G\varphi)$.
\end{enumerate}

Furthermore, over any class of serial frames (e.g., KD, KD45, and S5), conditions (4)--(7) are equivalent to:\footnote{For more equivalent conditions, see \citet[Theorems 13 and 14]{yamada2026Eventual}, including \emph{transfinite eventual self-refutation} and \emph{transfinite strong eventual self-refutation}.}
\begin{enumerate}[resume]
  \item $\varphi$ is commonly unbelievable.
  \item $\varphi$ is always informative.
\end{enumerate}
Over S5, conditions (1)--(9) are equivalent in either PAL or BPAL.
\end{theorem}
\begin{proof}
\emph{Over K: (1)--(3).}
Every transfinite orbit reaches a fixed point by Definition~\ref{def:stabilization-rank}.
Under (3), this terminal model is empty in PAL and edgeless in BPAL; in the latter case it retains the initial domain and valuation.
It therefore equals $U_\bot(M)$, proving (3)$\Rightarrow$(2).
Conversely, a fixed point is unchanged by every transfinite iterate, so applying (2) to a fixed point gives (3).
For finite $M$, Theorem~\ref{thm:stabilization-rank}(2) gives a finite stabilization stage, so (3) also implies (1).

To prove (1)$\Rightarrow$(3), suppose that $N$ is a fixed point with $N\ne U_\bot(N)$.
By Theorem~\ref{thm:fixed-points}(1), $N$ globally satisfies $\varphi$ in PAL and $\bigwedge_{i\in G}\Box_i\varphi$ in BPAL.
Let $\Sigma$ be the set of subformulas of the corresponding static formula, and take the minimal filtration $N_\Sigma$ through $\Sigma$.
Its states are the equivalence classes of agreement on $\Sigma$, and its relations are
\[
[x]R_i^\Sigma[y]
\quad\Longleftrightarrow\quad
x'R_i y'\text{ for some }x'\in[x],\ y'\in[y].
\]
The filtration truth lemma preserves all formulas in $\Sigma$, so $N_\Sigma$ is a fixed point with at most $2^{|\Sigma|}$ states.
In PAL it is nonempty, and in BPAL an original arrow induces an arrow between its equivalence classes.
Thus $N_\Sigma\ne U_\bot(N_\Sigma)$, contradicting (1).

\emph{(3)$\Leftrightarrow$(4) over K and S5.}

In PAL: We first see $(4)\Rightarrow (3)$ by contraposition. A nonempty fixed point globally satisfies $\varphi$, so each of its states satisfies $\varphi\land C_G\varphi$.
We next prove $(3)\Rightarrow (4)$ again by contraposition. If $M,w\models\varphi\land C_G\varphi$, then the generated submodel $M_w$ belongs to the class and globally satisfies $\varphi$, since static modal truth is preserved under generated submodels.
Thus $M_w$ is a nonempty PAL fixed point, proving (3)$\Rightarrow$(4).

In BPAL, every arrow target of a fixed point satisfies $\varphi$.
If a fixed point has an arrow $uR_iv$, then $v$ and every state positively reachable from $v$ satisfy $\varphi$.
Hence $\varphi\land C_G\varphi$ holds at $v$, proving (4)$\Rightarrow$(3).
Conversely, suppose $M,w\models\varphi\land C_G\varphi$, and put $N=M_w$, where $\varphi$ holds globally.
Over S5, $N$ is an S5 model and has an arrow, so it is a nonzero BPAL fixed point.
Over K, add a fresh state $z$ with the single outgoing arrow $zR_iw$, for some $i\in G$, and no arrows from old states to $z$.
Truth at old states is unchanged, so every arrow target still satisfies $\varphi$.
The enlarged model is a nonzero BPAL fixed point, proving (3)$\Rightarrow$(4).

\emph{(4)--(7) and (8)$\Leftrightarrow$(9).}
For any frame class, we have
$(M|\varphi)_w=M_w\iff M,w\models C_G\varphi$:
the generated submodel is unchanged exactly when no arrow in it is removed.
Thus (5) is equivalent to the validity of $\varphi\to\neg C_G\varphi$, which is also equivalent to (4), (6), and (7).
When $\varphi$ is true at $w$, the generated PAL and BPAL updates at $w$ coincide, giving the same reading of (5) for true PAL announcements.
Without a restriction on the truth value of $\varphi$, the same identity gives (8)$\Leftrightarrow$(9).

\emph{The serial classes: (4)$\Leftrightarrow$(8).}
If $M,w\models C_G\varphi$, seriality and $G\ne\varnothing$ give a successor $v$ of $w$.
Both $v$ and every state positively reachable from $v$ are positively reachable from $w$, so $M,v\models\varphi\land C_G\varphi$.
The converse implication between satisfiability of these two formulas is immediate.
Taking negations gives (4)$\Leftrightarrow$(8), completing the equivalence of (4)--(9) over any class of serial frames.

\emph{S5: (1)--(9).}
First consider PAL.
Every nonempty transfinite iterate of an S5 model is an induced S5 submodel, so (3)$\Rightarrow$(2) and (3)$\Rightarrow$(1) follow as above; (2)$\Rightarrow$(3) is unchanged.
For (1)$\Rightarrow$(3), suppose $N$ is a nonempty S5 fixed point.
Take its minimal filtration through $\Sigma=\operatorname{Sub}(\varphi)$ and replace each filtered relation by its transitive closure.
The minimal filtered relations are reflexive and symmetric, so their transitive closures are equivalence relations.
This remains a filtration: transitivity of the original $R_i$ propagates each $\Box_i\psi\in\Sigma$ along minimal filtered $i$-arrows, and reflexivity yields $\psi$ at the endpoint.
The truth lemma therefore gives a finite nonempty S5 model globally satisfying $\varphi$, hence a fixed point contradicting (1).
This proves (1)--(3) in PAL.

For BPAL, use the orbit correspondence~\eqref{eq:pal-bpal-orbit-correspondence}.
At every ordinal stage, the BPAL iterate is edgeless if and only if the PAL iterate is empty: each state in the PAL domain retains its original self-loops in BPAL.
Thus (1) and (2) have the same truth values in the two semantics over S5.
Together with the already established equivalence of (3)--(9) over S5, this proves the claim.
\end{proof}

The formulas constructed in the proof of Theorem~\ref{thm:stabilization-rank}(4) are always informative when true over K.
For $G=\{i\}$, if $\varphi=p\land\Diamond_i(p\land\Box_i\neg p)$ holds at $w$, its witnessing successor $v$ satisfies $p\land\Box_i\neg p$ and therefore falsifies $\varphi$.
For $|G|\geq2$, recall $\varphi=L\neg\delta$ with $\delta=L\top$; its witnessing successor satisfies $\neg\delta$ and hence $\neg\varphi$, since $\varphi\to\delta$ is valid.
Nevertheless, for each ordinal $\alpha>0$, the corresponding model $M_\alpha$ has $\widehat\rho_\varphi^P(M_\alpha)=\widehat\rho_\varphi^B(M_\alpha)=\alpha$.
Thus, being always informative when true does not imply fast stabilization of any model (i.e., there need not be a uniform bound on stabilization rank).
\section{Conclusion}\label{sec:conclusion}
In this paper, we introduced the public announcement operator $U_\varphi$ and studied its algebraic and dynamic properties with a particular focus on commutativity, monotonicity, idempotence, fixed points, and stabilization. 

There are several possible future directions. One direction is the decidability of finite stabilization: given a formula $\varphi$, decide whether $\rho_\varphi(M)<\omega$ for all models $M$. Another direction is to investigate whether some of our results fail for classes of frames other than K. Further directions include the epistemic interpretations of transfinite iterated announcements as noted in Section~\ref{subsec:stabilization}.

\backmatter

\section*{Statements and Declarations}
\bmhead{Acknowledgements}
I thank Koki Okura for his helpful comments during seminars. The author used GPT-6 Astra for reasoning, literature search, coding, drawing figures, and proofreading. Some of the proofs were written by this model but were thoroughly checked and modified by the author, who takes full responsibility for the final content.
\bmhead{Funding}
This research was supported by the Science Tokyo Support Program for Doctoral Students, funded by the Universities for International Research Excellence.
\bmhead{Competing interests}
The author has no competing interests to declare.
\bmhead{Author contributions}
The author is the sole author of the manuscript.
\bmhead{Data availability}
No datasets were generated or analyzed during the current study.
\bibliography{references}
\end{document}